\documentclass[twocolumn]{aastex62}

\usepackage{amsmath}
\usepackage{nicefrac}
\usepackage{xcolor}
\usepackage[normalem]{ulem}
\usepackage[applemac]{inputenc}

\defcitealias{SAGAII}{M21}

\newcommand\msun{\mathrm{M}_{\odot}}

\newcommand{\kms}{{\ensuremath{\mathrm{km\,s^{-1}}}}}
\newcommand{\SAGAII}{{SAGA\nobreakdash-II}}

\submitjournal{AAS Journals}

\shorttitle{Star-forming Satellites in Observations and Simulations}
\shortauthors{Karunakaran et al.}
\begin{document}

\title{Satellites Around Milky Way Analogs: Tension in the Number and Fraction of Quiescent Satellites Seen in Observations Versus Simulations}

\correspondingauthor{Ananthan Karunakaran}
\email{a.karunakaran@queensu.ca}

\author[0000-0001-8855-3635]{Ananthan Karunakaran}
\affil{Department of Physics, Engineering Physics and Astronomy, Queen's University, Kingston, ON K7L 3N6, Canada}
\author[0000-0002-0956-7949]{Kristine Spekkens}
\affil{Department of Physics and Space Science, Royal Military College of Canada P.O. Box 17000, Station Forces Kingston, ON K7K 7B4, Canada}
\affil{Department of Physics, Engineering Physics and Astronomy, Queen's University, Kingston, ON K7L 3N6, Canada}

\author[0000-0001-9857-7788]{Kyle A. Oman}
\affil{Institute for Computational Cosmology, Department of Physics, University of Durham, South Road, Durham DH1 3LE, UK}

\author[0000-0001-9985-1814]{Christine M. Simpson}
\affil{Enrico Fermi Institute, The University of Chicago, Chicago, IL 60637, USA}
\affil{Department of Astronomy and Astrophysics, University of Chicago, Chicago IL 60637, USA}

\author[0000-0002-6831-5215]{Azadeh Fattahi}
\affil{Institute for Computational Cosmology, Department of Physics, University of Durham, South Road, Durham DH1 3LE, UK}

\author[0000-0003-4102-380X]{David J. Sand}
\affil{Steward Observatory, University of Arizona, 933 North Cherry Avenue, Rm. N204, Tucson, AZ 85721-0065, USA}

\author[0000-0001-8354-7279]{Paul Bennet}
\affil{Space Telescope Science Institute, 3700 San Martin Drive, Baltimore, MD 21218, USA}

\author[0000-0002-1763-4128]{Denija Crnojevi\'{c}}
\affil{University of Tampa, 401 West Kennedy Boulevard, Tampa, FL 33606, USA}

\author[0000-0002-2338-716X]{Carlos S. Frenk}
\affil{Institute for Computational Cosmology, Department of Physics, University of Durham, South Road, Durham DH1 3LE, UK}

\author[0000-0002-1947-333X]{Facundo A. G\'{o}mez}
\affil{Instituto de Investigaci\'{o}n Multidisciplinar en Ciencia y Tecnolog\'{i}a, Universidad de La Serena, Ra\'{u}l Bitr\'{a}n 1305, La Serena, Chile}
\affil{Departamento de Astronom\'{i}a, Universidad de La Serena, Av. Juan Cisternas 1200 Norte, La Serena, Chile}

\author[0000-0001-9667-1340]{Robert J. J. Grand}
\affil{Max-Planck-Institut f\"{u}r Astrophysik, Karl-Schwarzschild-Str. 1, D-85748 Garching, Germany}

\author[0000-0002-5434-4904]{Michael G. Jones}
\affil{Steward Observatory, University of Arizona, 933 North Cherry Avenue, Rm. N204, Tucson, AZ 85721-0065, USA}

\author[0000-0003-3816-7028]{Federico Marinacci}
\affil{Department of Physics \& Astronomy ``Augusto Righi'', University of Bologna, via Gobetti 93/2, I-40129 Bologna, Italy}

\author[0000-0001-9649-4815]{\textsc{Bur\c{c}{\rlap{\.}\i}n} \textsc{Mutlu-Pakd{\rlap{\.}\i}l}}
\affil{Kavli Institute for Cosmological Physics, University of Chicago, Chicago, IL 60637, USA}
\affil{Department of Astronomy and Astrophysics, University of Chicago, Chicago IL 60637, USA}

\author[0000-0003-3862-5076]{Julio F. Navarro}
\affil{Department of Physics and Astronomy, University of Victoria, P.O. Box 3055, STN CSC, Victoria BC V8W 3P6, Canada}

\author[0000-0002-5177-727X]{Dennis Zaritsky}
\affil{Steward Observatory, University of Arizona, 933 North Cherry Avenue, Rm. N204, Tucson, AZ 85721-0065, USA}

\begin{abstract}We compare the star-forming properties of satellites around Milky Way (MW) analogs from the Stage~II release of the Satellites Around Galactic Analogs Survey (SAGA-II) to those from the APOSTLE and Auriga cosmological zoom-in simulation suites.\ We use archival {\it GALEX} UV imaging as a star-formation indicator for the \SAGAII{} sample and derive star-formation rates (SFRs) to compare with those from APOSTLE and Auriga.\ We compare our detection rates from the NUV and FUV bands to the \SAGAII{} H$\alpha$ detections and find that they are broadly consistent with over $85\%$ of observed satellites detected in all three tracers.\ We apply the same spatial selection criteria used around \SAGAII{} hosts to select satellites around the MW-like hosts in APOSTLE and Auriga.\ We find very good overall agreement in the derived SFRs for the star-forming satellites as well as the number of star-forming satellites per host in observed and simulated samples.\ However, the number and fraction of quenched satellites in the \SAGAII{} sample are significantly lower than those in APOSTLE and Auriga below a stellar mass of $M_*\sim10^{8}\,\msun$, even when the \SAGAII{} incompleteness and interloper corrections are included.\ This discrepancy is robust with respect to the resolution of the simulations and persists when alternative star-formation tracers are employed.\ We posit that this disagreement is not readily explained by vagaries in the observed or simulated samples considered here, suggesting a genuine discrepancy that may inform the physics of satellite populations around MW analogs. \end{abstract}

\keywords{Dwarf galaxies (416), Galaxy evolution (594), Galaxy quenching (2040), Quenched galaxies (2016), Star formation (1569)
\vspace{.1cm}}
\section{Introduction}\label{sec:intro}
Characterizing the satellite populations around Milky Way--like hosts is a key component of understanding galaxy formation and evolution.\ Owing to the unrivaled depth and completeness of observations of its satellite population, the Milky Way (MW) has been the default test-bed for simulations aiming to probe the underlying physics of dwarf galaxy evolution.\ The environmental dependence of the satellite star forming fraction therein is well-established: with few massive ($M_* \gtrsim 10^8 \msun$) exceptions, satellites are quiescent within the virial radius and star forming farther out \citep[][]{HILocalGroup,Spekkens2014,2021Putman}, implying that satellites are quenched by their hosts.\ Indeed, it has been demonstrated that the observed LG satellite quenched fraction transitions from $\sim100\% - 0\%$ between $10^{6.5} \lesssim M_{*}/\msun \lesssim 10^{9}$, suggesting a mass dependence to the underlying mechanisms (\citealt[][]{2015fillingham,2015Wetzel}, see also \citealt[][]{2014Slater,2014Wheeler}).\ Whether or not the star-forming properties of satellites around MW-mass hosts beyond the Local Group (LG) are consistent with those within the LG can provide important constraints on cosmological galaxy formation simulations.\

Observationally, searches for the satellite populations around nearby MW anologs were pioneered by \citet[][]{1993Zaritsky,1997Zaritsky} and are now being pursued by various groups \citep[e.g.][]{2016Crnojevic,2016Javanmardi,2019Bennet,2020Carlsten}.\ One of the most extensive of these campaigns is the ongoing Satellites Around Galactic Analogs \citep[SAGA,][]{SAGAI} survey, which aims to detect and characterize all satellites brighter than the Leo I dwarf ($M_*\sim10^{6.6}\msun$) around 100 MW-like hosts.\ The SAGA Stage II (\SAGAII{}) release presented in \citet[][hereafter M21]{SAGAII} shows that the vast majority of confirmed satellites within the virial radii of the 36 hosts surveyed so far are star forming rather than quenched.\ While this result is commensurate with some earlier surveys of brighter satellites across a broader host mass range \citep[e.g.][]{2013Guo,2015Phillips,2019Davies}, it is in strong contrast to the fainter satellites in the LG.\ This discrepancy between the LG and other observed systems may have important implications for models that aim to replicate the trends seen in the LG.\

Theoretically, a variety of cosmological zoom-in simulation suites can now probe star formation and quenching physics in satellites around MW analog hosts down to Leo I masses \citep[][]{2016Wetzel,2019GarrisonKimmel,Akins2021}.\ In particular, galaxy properties in APOSTLE \citep[][]{2016Sawala,2016Fattahi} and Auriga \citep[][]{2017Grand} are interesting to contrast given their similar resolutions and suite sizes but different host selection, hydrodynamical schemes (SPH vs.\ moving-mesh), and evolutionary models.\ One focus has been the comparison of the simulated satellite quenched fraction to that in the LG.\ The agreement between simulations is generally good, with most studies suggesting a characteristic mass ($M_{*} \sim10^{8-9}\msun$) below which satellites are more readily quenched by their hosts \citep[e.g.][]{2016Fillingham,2018Simpson,Akins2021,2021Joshi}.\ 

The consistency of simulated satellite populations with those in the LG combined with the stark contrast between the quenched fractions in the MW and \SAGAII{} strongly motivate direct comparisons between theory and other MW analogs in order to build robust models of galaxy formation.\ This requires selecting star-forming objects consistently across observed and simulated samples.\

In this letter, we compare star-forming satellites and quenched fractions in the \SAGAII{} observations to those in the APOSTLE and Auriga simulations.\ The observations and simulations have similar host numbers, host masses, and satellite selection functions (\S\ref{sec:sample}).\ We use archival UV imaging and simulated star formation rates to select star-forming satellites in \SAGAII{} and APOSTLE/Auriga, respectively, comparing star formation rates (SFRs) to gauge consistency across samples (\S\ref{sec:Obsanddata}).\ We then compare quenched fractions in the observed and simulated samples (\S\ref{sec:results}) and discuss possible explanations for the significant discrepancies we find (\S\ref{sec:Conclusion}).\

\section{Satellite Samples} \label{sec:sample}

\subsection{{Observed sample: SAGA-II}} \label{subsec:saga}

We adopt the ``complete systems" in the \SAGAII{} release as our observed sample, which consists of 127 confirmed satellites across 36 surveyed hosts with $M_{\mathrm{halo}}\sim (0.7-2)\times10^{12}\,\msun$\footnote{Estimated in \SAGAII{} following \citet[][]{2020Nadler} where virial parameters are estimated at $\simeq99.2$ times the critical density of the Universe, $\rho_{crit}$.}.\ \SAGAII{} hosts are selected primarily on luminosity $(-23>M_K >-24.6)$, are largely in the field with a few that are members of LG-like pairs (see \citetalias{SAGAII} for details), and are mostly star-forming galaxies.\ We use the \SAGAII{} optical properties, stellar masses and distances derived for all observed hosts and satellites, which are reproduced in Table~\ref{table:maintable}.\

As explained in detail in \citetalias{SAGAII}, imaging catalogs are used to build satellite candidate lists around each host, and candidates without archival redshifts are targeted spectroscopically to confirm an association.\ Sample-wide, 80$\%$ (100$\%$) of candidates with extinction-corrected (designated with subscript ``o") $r$-band absolute magnitudes $M_{r,o} \leq -12.3$ ($-15.5$) are targeted spectroscopically in \SAGAII{}.\ We convert these limits to stellar masses $\mathrm{log}\big[\frac{M_*}{\msun}\big] \sim 6.6$ \big($\mathrm{log}\big[\frac{M_*}{\msun}\big] \sim 7.8$\big) using the relations in \citetalias{SAGAII} and an average satellite sample color of $(g-r)_o \sim 0.39$.\ Since star-forming satellites are easier to detect than quiescent ones in H$\alpha$ the spectroscopic coverage is only indirectly related to completeness, particularly for quiescent systems.\ \citetalias{SAGAII} undertake detailed modelling to estimate the impact of incompleteness and interlopers on the sample quenched fraction, which we adopt here (see \S\ref{sec:results}).

The single-fibre H$\alpha$ measurements in \SAGAII{} provide an estimate of star formation activity that is not amenable to direct comparisons with simulations \citepalias{SAGAII}.\ We therefore make use of data from the {\it Galaxy Evolution Explorer} \citep[{\it GALEX},][]{2005MartinGALEX} to search for UV emission in \SAGAII{} satellites to provide a homogeneous, global star formation activity indicator for each system.\ In total, 119/127 \SAGAII{} satellites have archival NUV and/or FUV coverage, and we select the deepest available imaging for the search (26/58 tiles have depths greater than \textit{GALEX} AIS data, i.e.\ integration times $\gg300$ seconds; Table~\ref{table:maintable}).\

\begin{figure*}[htb!]
\centering
\includegraphics[width=13.5cm]{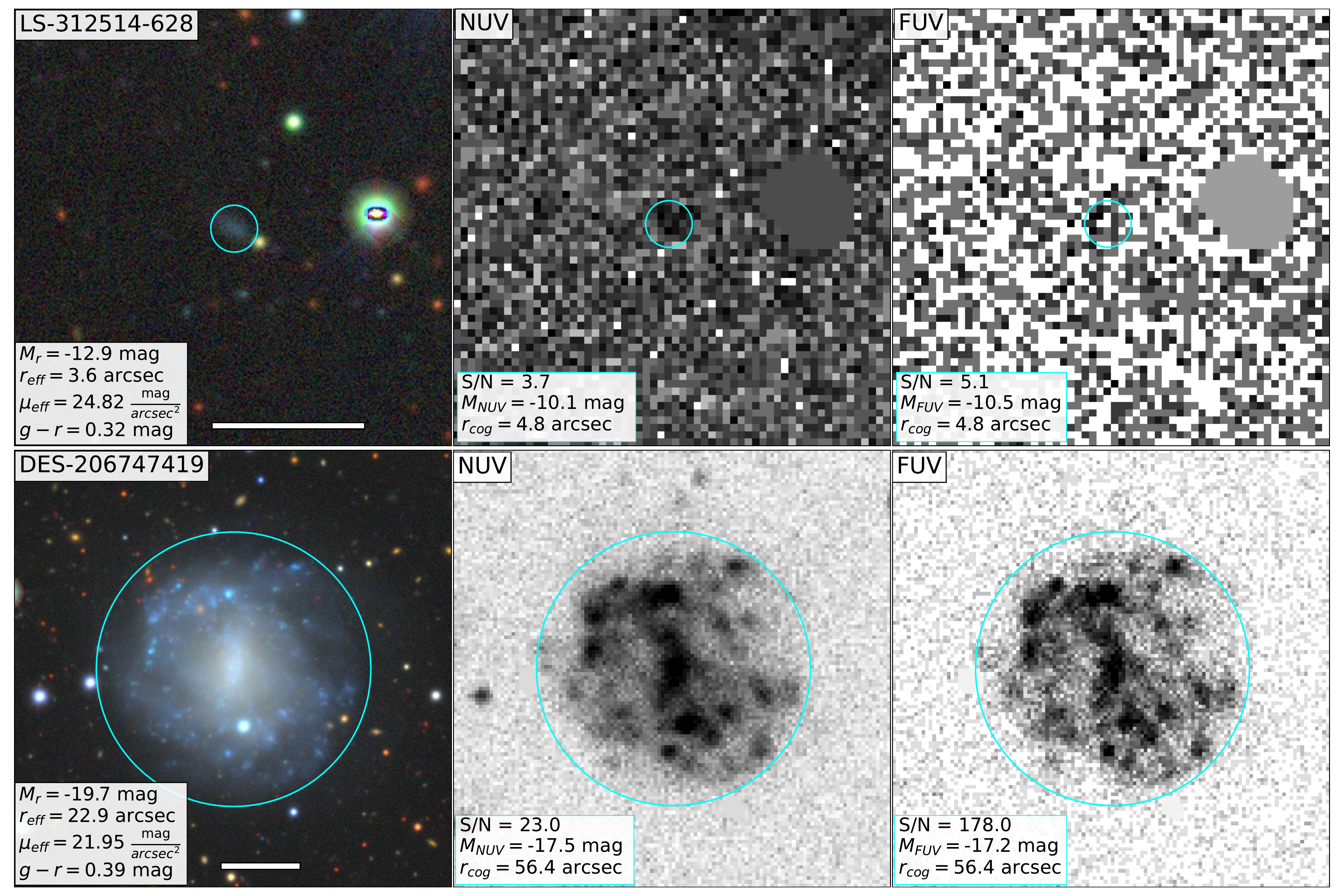}
\caption{Optical (left, composite $grz$ from the DESI Legacy Surveys Imaging DR9; \citealt{2019Dey}), masked NUV (middle), and masked FUV (right) image cutouts of a small, faint (top) and large, bright (bottom) observed satellite to illustrate our curve-of-growth UV measurement method.\ The cyan circles represent the aperture with radius $r_{\mathrm{COG}}$, within which the UV emission is measured.\ Optical properties from \citetalias{SAGAII} and UV properties that we measure are shown in the bottom-left of the corresponding panel.\ The scale bar at the bottom of the left panels represents 30 arcseconds.\ }
\label{fig:cogexample}
\end{figure*}

\subsection{{Simulated samples: APOSTLE and Auriga}} \label{subsec:sims}
We adopt hosts and satellites from APOSTLE \citep[][]{2016Sawala,2016Fattahi} and Auriga \citep[][]{2017Grand} to define simulated samples.\ The APOSTLE suite traces the formation and evolution of LG-like environments with MW-M31 pairs (selected by halo mass, separation, and kinematics) and their surrounding environment.\ In contrast, the Auriga project simulates isolated MW-like halos.\ Both suites invoke differing models for galaxy formation and evolution which include prescriptions for all relevant physical processes (i.e.\ gas cooling, stellar and AGN feedback, UV background, etc.).\ For more details on the EAGLE model used in APOSTLE, see \citet[][]{2015Schaye} and \citet[][]{2015Crain}; for Auriga details, see \citet[][]{2017Grand}.\ 

For APOSTLE, we consider the 12 intermediate-resolution (L2) MW-M31 analog pairs for a total of 24 distinct satellite systems around hosts with $M_{\mathrm{halo}}\sim(0.5-2.4)\times10^{12}\,\msun$\footnote{Halo masses in both APOSTLE and Auriga are calculated within the radius that encompasses a mean matter density equal to 200 times $\rho_{crit}$.\ }.\ For Auriga, we consider the satellite systems of 37 standard resolution (Level 4) non-merging hosts with $M_{\mathrm{halo}}\sim(0.4-2)\times10^{12}\,\msun$ \citep[][]{2018Simpson,2019Grand}.\ The adopted simulations have comparable dark matter particle ($m_{\mathrm{DM}}\sim5.9\times 10^5\,\msun\mathrm{\,vs.\,} \sim 3\times 10^5\,\msun$) and stellar/baryon ($m_{\mathrm{star}}\sim 1.2\times 10^5\,\msun\mathrm{\,vs.\,}5\times10^4\,\msun$) resolutions, respectively.\ We test convergence with higher resolution volumes available for both APOSTLE and Auriga in Appendix~\ref{sec:restests} and find no significant deviation in the estimated satellite quenched fractions from the standard resolution volumes.\ 

We define the simulated satellite population by selecting from the set of SUBFIND \citep[][]{2001Springel} subhalos that have embedded galaxies with stellar masses within two spherical stellar half-mass radii of $M_{*} \geq 10^6\,\msun$ and are within an aperture of radius $400\,$kpc around each host.\ We note that we tested smaller and larger spherical apertures (i.e.\ 300 kpc and 1 Mpc) around Auriga hosts and find a minimal difference ($<5\%$) on our final results, likely due to the application of the \SAGAII{} selection function (see below).\ 

We take a single random projection (different projections produce nearly identical results on the whole) for each host to define the sample, although we orient the line connecting APOSTLE host pairs away from the line of sight to minimize the effects of interlopers from the other host.\ For a given simulation volume orientation, the sample selection criteria mimic those of \SAGAII{}: we choose the set of these subhalos with projected separations $10\,\mathrm{kpc} \leq D_{\mathrm{proj}} \leq 300\,\mathrm{kpc}$ and relative line-of-sight velocities $|\Delta V_{\mathrm{sys}}| \leq 275\,\kms$ of their host.\ This produces a simulated APOSTLE sample of 229 satellites, and a simulated Auriga sample of 411 satellites.\ We discuss the similarities and differences between these simulated samples in \S\ref{subsec:simSF}.\ 

\section{Identifying Star-Forming Satellites} \label{sec:Obsanddata}

With the satellite samples established in \S \ref{sec:sample}, we now outline our method to select observed (\S \ref{subsec:obsSF}) and simulated (\S \ref{subsec:simSF}) star-forming satellites within them.\ We check for consistency of our star-forming satellite definitions across the observations and the simulations in \S \ref{subsec:compSF}.

\subsection{Observed Star-Forming Satellites}
\label{subsec:obsSF}

We use UV emission as the primary indicator of star formation in the observed satellites from \SAGAII{} with archival {\it GALEX} imaging (see Table~\ref{table:maintable}).\ We take a curve-of-growth approach using the Astropy Photutils package \citep[][]{larrybradley2020} to detect statistically significant UV emission.\ Our method is illustrated in Figure~\ref{fig:cogexample}, and the corresponding measurements are in Table~\ref{table:maintable}.\

We start by masking bright sources near the satellite targets in each 1.2 degree-wide {\it GALEX} tile and measuring the mean and standard deviation of the flux within 1000 randomly-placed circular regions across them.\ The region radius is the satellite effective radius $r_{\mathrm{eff}}$ from the \SAGAII{} photometry.\ Working from the (generally deeper) NUV tile, we measure background-subtracted fluxes within circular apertures about the optical position of each satellite starting from $r=0.5 r_{\mathrm{eff}}$.\ We increase the aperture size in steps of $0.75$\arcsec ($3$\arcsec) for less (more) extended sources until the background-subtracted fluxes in adjacent apertures change by less than the noise difference between them.\ We compute the signal-to-noise $S/N_{{NUV}}$ in the smaller of these regions (with a radius $r_{\mathrm{COG}}$ reported in Table~\ref{table:maintable}), and place an identical region on the FUV tile to measure $S/N_{{FUV}}$.\

We consider measured fluxes with $S/N > 2$ as detections in a given band.\ By this definition, 115/119 satellites with {\it GALEX} coverage have associated NUV emission, and 104/113 have associated FUV emission.\ We use standard equations \citep[][]{2007MorrisseyGALEX} to convert to apparent AB magnitudes $m_{NUV}$ and $m_{FUV}$ (see Table~\ref{table:maintable}), correcting for foreground extinction using $E(B-V)$ from \citet{2011Schlaflydust} with $R_{NUV} = 8.2, R_{FUV} = 8.24$ \citep[][]{2007wyder}. 

 Not only do the vast majority of the \SAGAII{} satellites with UV coverage show emission in one or both bands, but the correspondence between satellites with H$\alpha$ equivalent widths $\mathrm{EW} \geq 2\mathrm{\AA}$ (\citetalias{SAGAII}) is also very high: 98/113 satellites with observations in NUV, FUV and H$\alpha$ are detected in all three tracers.\ We posit that the majority of non-detections stem from observational limitations (such as image depth/sensitivity combined with satellite distances or H$\alpha$ fiber position) rather than physical differences.\ The strong correlation between UV and H$\alpha$ star formation tracers, despite the difference in the timescales they probe, is common in dwarf galaxies \citep[e.g.][]{2011Lee}, and suggests that quenching is rapid at these masses \citep[e.g.][]{2015Wetzel}.\ 
 
 We define an observed satellite to be star forming either if it is detected in the UV or if it has $\mathrm{EW} \geq 2\mathrm{\AA}$ as reported by \citetalias{SAGAII}.\ Since we find 12 (6) satellites with NUV (FUV) emission that do not satisfy the H$\alpha$ criterion but only 2 satellites (1 NUV, 1 FUV) for which the inverse is true, the fraction of star-forming satellites in Table~\ref{table:maintable}, 120/127, is marginally higher than that reported by \citetalias{SAGAII}.\ We discuss the implications of these numbers for the quenched fraction in \S \ref{sec:results}.
 
\begin{figure*}[htb!]
\centering
\includegraphics[width=\columnwidth]{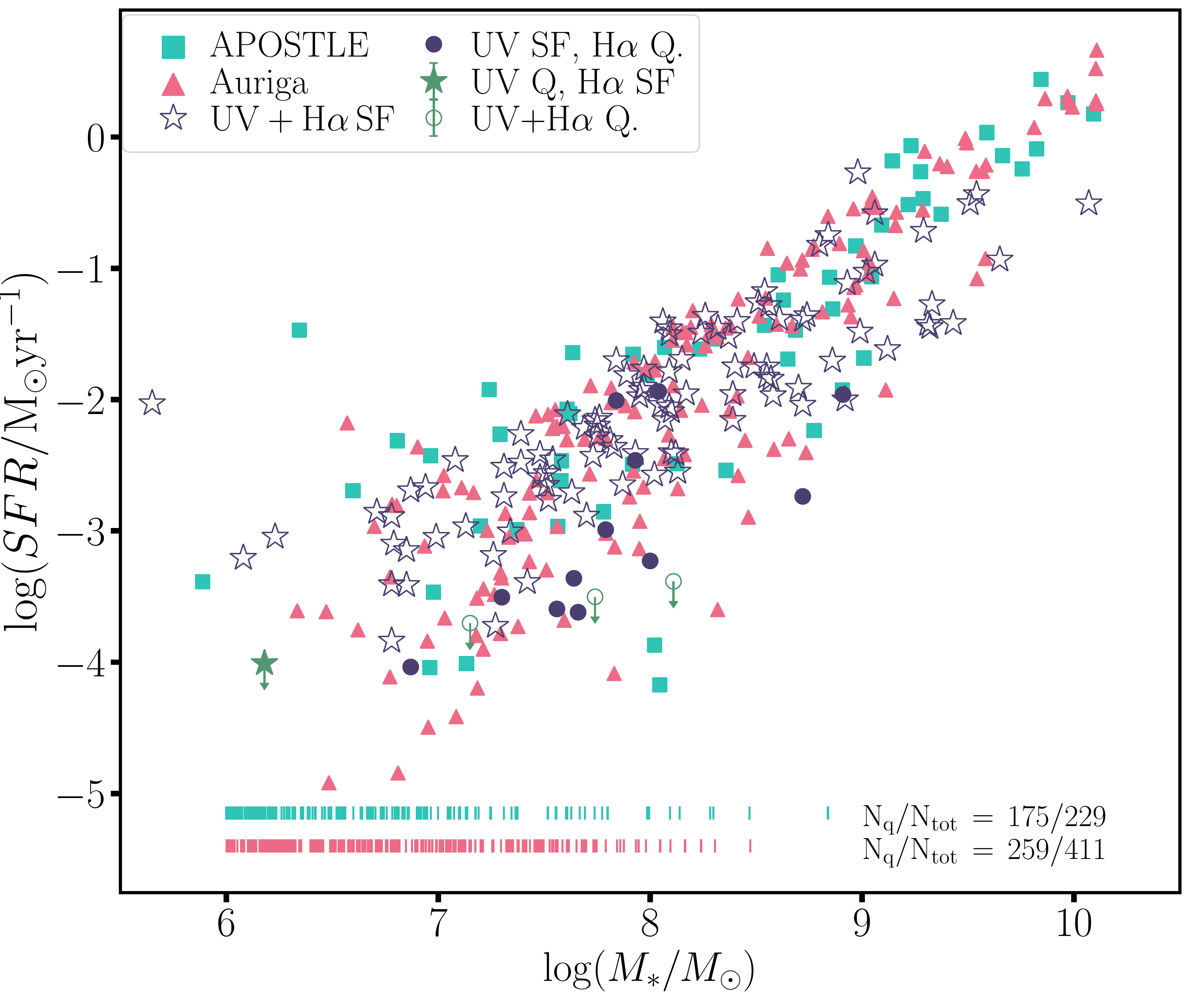}
\includegraphics[width=\columnwidth]{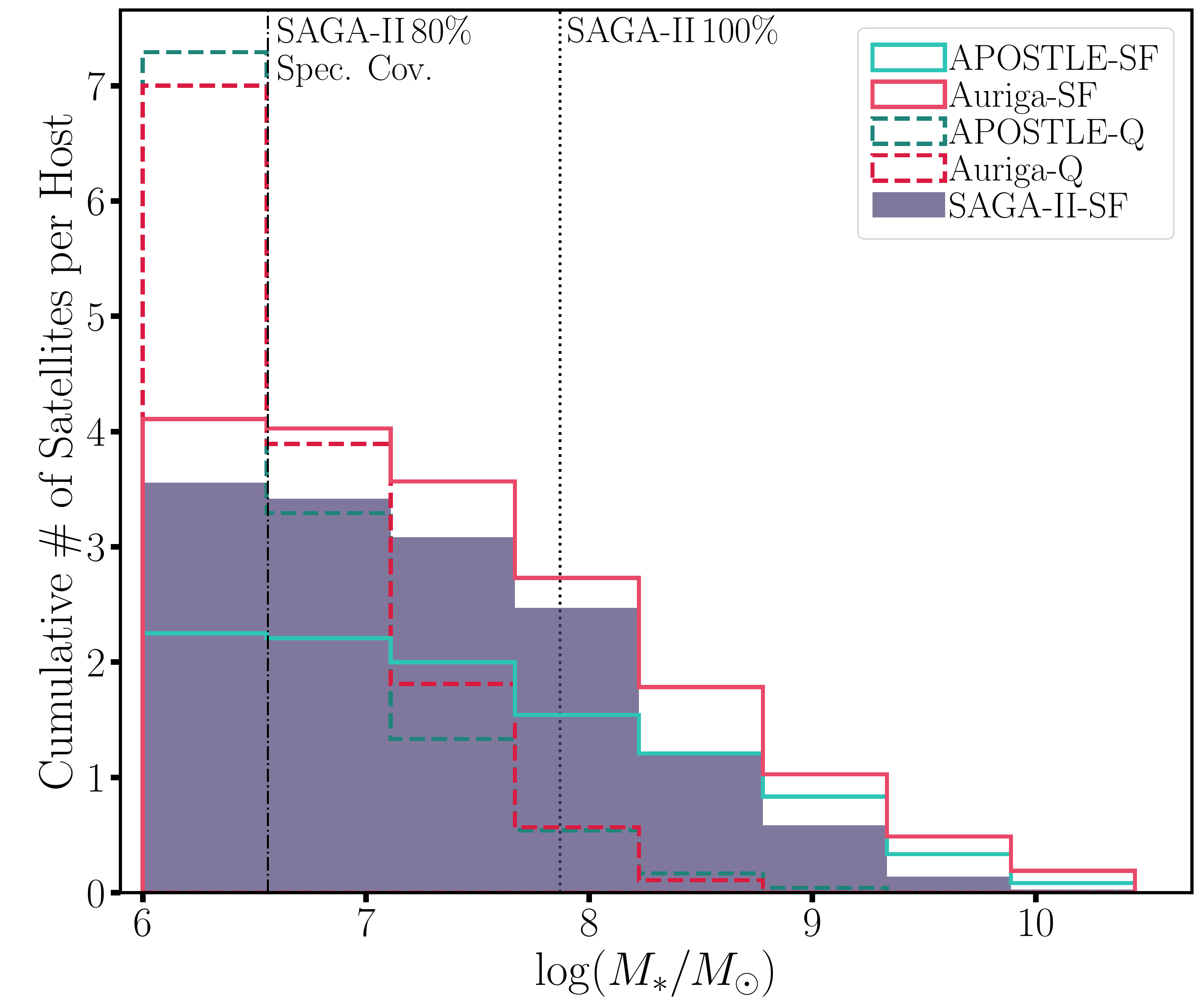}
\caption{{\it Left:} Simulated satellite $\mathrm{SFR_{sim}}$-$M_{*}$ relation derived from APOSTLE (cyan squares) and Auriga (pink triangles).\ Quenched simulated satellites are represented as short vertical lines at their $M_{*}$, along with quenched/total number counts $\mathrm{N_q/N_{tot}}$ to the right.\ The SFR$_{NUV}$-$M_{*}$ relation for observed satellites is overplotted in green and purple, with the symbol shape and color indicating whether or not the satellite is star-forming (SF) from UV and/or H$\alpha$ tracers as explained in the legend.\ {\it Right:} Cumulative number of satellites per host for the three samples.\ The quenched and star-forming satellites from the two simulations are shown as dashed and solid lines, respectively.\ The solid purple histogram shows star-forming satellites from \SAGAII{}, with the vertical dotted (dash-dotted) lines showing 100\% (80\%) spectroscopic coverage.\ There is very good agreement between the observed and simulated star-forming satellites by these metrics.\ There is also a population of low-mass quenched satellites in the simulations that has no counterpart in the observed satellite list in Table~\ref{table:maintable}.}
\label{fig:sfr-stellar}
\end{figure*}

\subsection{Simulated Star-Forming Satellites}
\label{subsec:simSF}
We consider two SFR measures to identify star-forming satellites in the simulations.\ Our fiducial metric, the ``instantaneous" rate, $\mathrm{SFR}_{\mathrm{sim}}$, is estimated using the gas particles associated with the satellite subhalo determined by SUBFIND at $z=0$ and corresponding star-formation rate relations for APOSTLE \citep[][]{2015Schaye} and Auriga \citep[][]{2003SpringelHernquist,2017Grand}, with the former using a gas pressure threshold and the latter using a gas density threshold.\ These SFRs have previously been shown to reproduce observed trends \citep[][]{2013Vogelsberger,2015Furlong,2015Schaye}.\ We also consider the average mass of star particles formed over the last gigayear as a measure of SFR.\ Like $\mathrm{SFR}_{\mathrm{sim}}$ this metric is less susceptible to shot noise than estimates over shorter time intervals, but, unlike the observational tracers, it averages over a significant fraction of a satellite orbit.\ We demonstrate in Appendix~\ref{sec:restests} that both metrics produce similar results, and adopt $\mathrm{SFR}_{\mathrm{sim}}$ to select simulated star-forming satellites throughout.

We define a simulated satellite as star-forming if $\mathrm{SFR}_{\mathrm{sim}} > 0 \, \msun\mathrm{yr^{-1}}$.\ A total of 54/229 APOSTLE and 152/411 Auriga satellites meet this criterion, and their properties are illustrated in Figure~\ref{fig:sfr-stellar}.\ The left-hand panel of Figure \ref{fig:sfr-stellar} shows the $\mathrm{SFR}_{\mathrm{sim}} - M_{*}$ relation of star-forming satellites and the stellar mass distribution of the quenched ones.\ The right-hand panel of Figure \ref{fig:sfr-stellar} shows the cumulative number of star forming and quenched satellites per simulated host.\ Considering the differences between the simulations (see \S \ref{subsec:sims}), there is good agreement between them despite the different host environments: star-forming satellites in both APOSTLE and Auriga follow similar $\mathrm{SFR}_{\mathrm{sim}} - M_{*}$ relations, and both populations become increasingly dominated by quenched systems at low $M_*$.\ These trends are qualitatively similar if a specific SFR threshold is adopted instead of a non-zero $\mathrm{SFR}_{\mathrm{sim}}$ \citep[i.e.][]{Akins2021}.

The mild difference between the APOSTLE and Auriga satellite samples in the right panel of Figure \ref{fig:sfr-stellar} likely stems from the different galaxy formation prescriptions adopted by the simulations.\ For example, APOSTLE halos may more readily remove gas from lower-mass subhalos leaving them permanently quenched, while Auriga subhalos may re-accrete expelled gas allowing them to be more long-lived.\ Additionally, the earlier onset of the UV background in the APOSTLE simulations may further contribute to fewer star-forming satellites per host at lower $M_*$.\

\subsection{Comparing Star-Forming Satellites}
\label{subsec:compSF}

To check for consistency between our definition of star-forming satellites in the observed and simulated samples, we estimate SFRs for the \SAGAII{} satellites with NUV detections, $\mathrm{SFR_{obs}}$, to compare with $\mathrm{SFR_{sim}}$ for simulated objects.
We use Equation~3 from \citet{2006IglesiasParamo} and the NUV luminosity, $L_{NUV}$, calculated from $m_{NUV}$ assuming the satellite to be at the distance of its host (see Table \ref{table:maintable}).\ We do not perform internal extinction corrections to $L_{NUV}$ in estimating SFRs, since homogeneously-measured infrared (IR) fluxes would be required and since our main interest in the SFRs themselves is diagnostic.\ The IR correction is likely small at the low-$M_*$ end of the satellite distribution, but significant (and uncertain) at higher masses \citep[e.g.][]{2015McQuinn}.\ The values of $\mathrm{SFR_{obs}}$ in Table~\ref{table:maintable} are therefore approximate, and likely represent lower limits at the high-mass end.\ 

We check for consistency between $\mathrm{SFR_{obs}}$ and $\mathrm{SFR_{sim}}$ in the left panel of Figure \ref{fig:sfr-stellar}, where the $\mathrm{SFR_{obs}} - M_{*}$ relation for the observed sample is plotted with the symbol shapes and colors identifying star-forming satellites according to UV and/or H$\alpha$ criteria.\ It is clear that there is broad agreement between the observed and simulated star-forming satellites\footnote{The object with the lowest $M_{*}$ in the observed sample (LS-330948-4542) appears to have a size that is severely under-estimated in the \citetalias{SAGAII} catalog, which likely explains its outlying $\mathrm{SFR_{obs}}$ in the left panel of Figure \ref{fig:sfr-stellar}.}, with the lower $\mathrm{SFR_{obs}}$ at $M_* \gtrsim 10^9 \, \msun$ relative to $\mathrm{SFR_{sim}}$ likely stemming from the lack of an IR correction in the former.\ The cumulative distribution of observed star-forming satellites per host, shown in the right panel of Figure \ref{fig:sfr-stellar}, also compares favourably to that from the simulations, with the observed distribution falling in between those for APOSTLE and Auriga for $M_* \lesssim 10^{8.5}\, \msun$ and slightly below both of them for $M_* \gtrsim 10^{9}\, \msun$.\ It is also interesting to briefly consider the effect of the hosts in this comparison.\ The majority of hosts in the observed sample and all of those in the simulated sample are star-forming galaxies.\ Both samples also demonstrate the concept of 'galactic conformity', where the properties of the satellites match those of their hosts \citep[e.g.][]{2014Phillips}, at higher satellite masses.\

Taken as a whole, Figure \ref{fig:sfr-stellar} suggests broad consistency between the definition of a star-forming satellite in the observed sample and in the simulated samples.\ It is also clear from the paucity of open green circles relative to the short vertical lines in the left panel of Figure \ref{fig:sfr-stellar} that there is a population of low-mass quenched satellites in the simulations that has no counterpart in the observed satellite list in Table~\ref{table:maintable}; the right panel of Figure \ref{fig:sfr-stellar} illustrates that a significant fraction of the quenched simulated satellites fall within the 80\% -- 100\% spectroscopic coverage limits for \SAGAII{}.\ This suggests a higher number and fraction of quenched satellites in the simulated samples than in the observed one, although incompleteness and interlopers in the latter need to be considered.\ We compare quenched fractions in the next section.

\section{Observed and Simulated Quenched Fractions} \label{sec:results}

With star-forming satellites identified and their consistency checked, we proceed to compare observed and simulated quenched fractions.

Because most \SAGAII{} candidates are confirmed in the H$\alpha$ emission line (see \S \ref{subsec:saga}), Table~\ref{table:maintable} is likely missing quenched satellites even in regions where the spectroscopic coverage is high.\ Interloping field galaxies are also more likely to be star-forming than quenched given their relative ubiquity \citep[][]{2012Geha}.\ Correcting for both effects would systematically raise the observed quenched fraction relative to that calculated directly from Table~\ref{table:maintable}.\ \citetalias{SAGAII} model them in detail, deriving a (dominant) incompleteness correction by assuming that all spectroscopically targeted but undetected candidates are quenched satellites, and a (subdominant) interloper correction by drawing mock samples from gravity-only simulations.\ We use the \citetalias{SAGAII} corrections directly from their Figure 11 (in the same $M_*$ bins), retaining the interloper correction despite comparing to simulations (for which it should not be required).\ The incompleteness/interloper-corrected quenched fractions we adopt are therefore conservative upper limits on the observed values implied by \SAGAII{}.

\begin{figure}[htb!]
\includegraphics[width=\columnwidth]{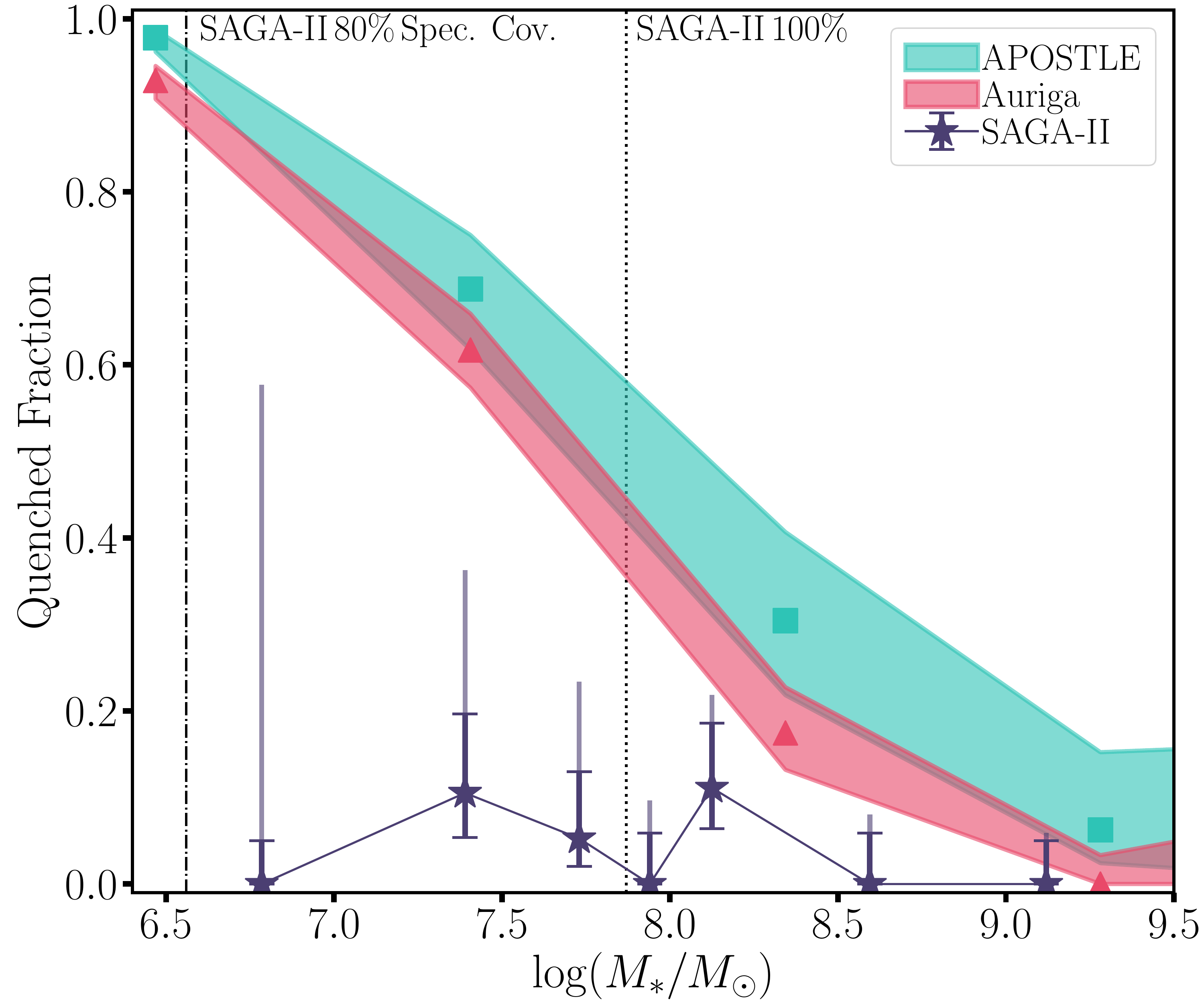}
\caption{Satellite quenched fractions as a function of stellar mass.\ The dark purple stars show the observed quenched fraction, with dark purple bars showing random uncertainties at 68\% confidence and the light purple bars showing the systematic incompleteness and interloper corrections from \citetalias{SAGAII} (see text).\ The dotted (dash-dotted) lines show the \SAGAII{} 100\% (80\%) spectroscopic coverage.\ The pink triangles, cyan squares and corresponding colored bands show the simulated quenched fraction and 68\% confidence intervals from APOSTLE and Auriga, respectively.\ Even accounting for incompleteness, there is a clear discrepancy between the observed and simulated quenched fractions for $10^7 \lesssim M_*/\msun \lesssim 10^8$, where the \SAGAII{} spectroscopic coverage is high. }

\label{fig:quenchfracsfhist}
\end{figure}

\begin{figure*}[htb!]
\includegraphics[width=18cm]{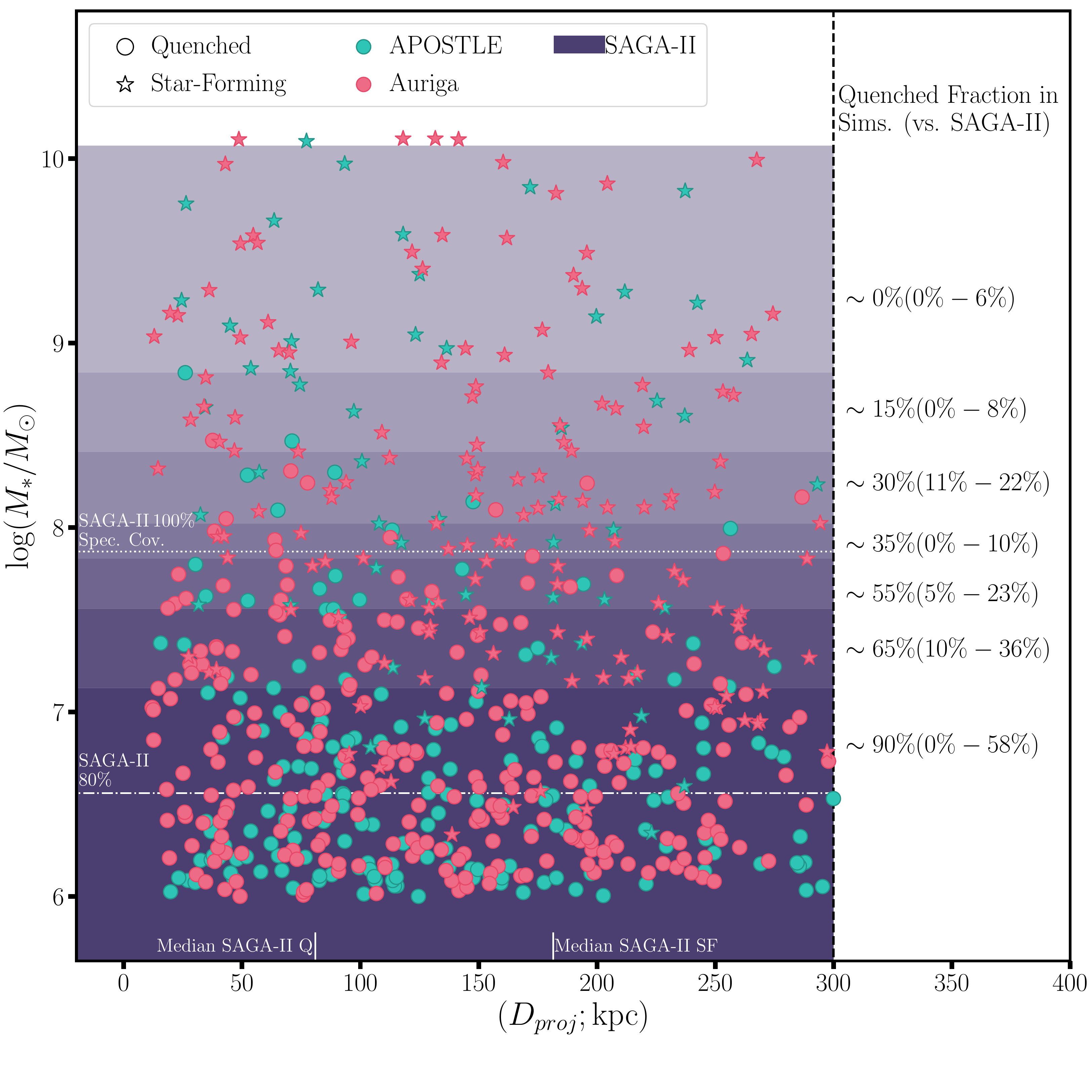}
\caption{Stellar mass as a function of projected distance $D_{\mathrm{proj}}$ of star-forming (stars) and quenched (circles) simulated satellites from APOSTLE (cyan) and Auriga (pink).\ The horizontal purple bands show the observed quenched fraction bins from \citetalias{SAGAII} and Figure \ref{fig:quenchfracsfhist}.\ The average simulated quenched fraction in each band is given to the right of it, and the numbers in parentheses show the measured (smaller) and incompleteness/interloper-corrected (larger) quenched fractions in the observed sample.\ The horizontal dotted (dash-dotted) lines show the \SAGAII{} 100\% (80\%) spectroscopic coverage, and the short vertical bars at the bottom show the median quenched (left) and star-forming (right) of the observed satellites in Table~\ref{table:maintable}.\ There are hints of a $D_{\mathrm{proj}}$ dependence of the quenched fraction in the simulations and observations.\ Even accounting for incompleteness, there is a discrepancy between the observed and simulated quenched fractions for $10^7 \lesssim M_*/\msun \lesssim 10^8$, where the \SAGAII{} spectroscopic coverage is high. \vspace{2cm} }

\label{fig:satcomp}
\end{figure*}

Figures~\ref{fig:quenchfracsfhist}~and~\ref{fig:satcomp} plot the observed and simulated quenched fractions in two different ways.\ In Figure \ref{fig:quenchfracsfhist}, the \SAGAII{} quenched fractions (purple stars with dark bars showing random counting uncertainties\footnote{68\% confidence intervals calculated using the Wilson score interval \citep[][]{brown2001interval} for both the observed and simulated samples.\ } and light bars showing systematic incompleteness/interloper corrections) are compared to those in APOSTLE (cyan band and squares) and Auriga (pink band and triangles).\ Figure \ref{fig:satcomp} plots $M_*$ as a function of projected host distance $D_{\mathrm{proj}}$ for star-forming (stars) and quenched (circles) satellites in the APOSTLE (cyan) and Auriga (pink) samples.\ The $M_*$ bin definitions in \citetalias{SAGAII} and Figure \ref{fig:quenchfracsfhist} are shown as a gradient of purple horizontal bands in Figure \ref{fig:satcomp}.\ The average simulated quenched fraction in those bins is directly to their right, and the range of observed quenched fractions bracketed by the ratios from Table~\ref{table:maintable} (smaller value) and the incompleteness/interloper-corrected ratios (larger value) are in parentheses.\ In both plots, the dotted (dash-dotted) lines show the \SAGAII{} 100\% (80\%) spectroscopic coverage.\

Figures \ref{fig:quenchfracsfhist}~and~\ref{fig:satcomp} illustrate that, despite their differences (i.e.\ hydrodynamic schemes, galaxy formation and evolution models, and host environments), there is a striking correspondence between the quenched fractions from the APOSTLE and Auriga simulations as a function of $M_*$ and $D_{\mathrm{proj}}$ for $10^6 \lesssim M_*/\msun \lesssim 10^{10}$.\ It is also clear that, even when conservatively accounting for both incompleteness and interlopers as in \citetalias{SAGAII}, the observed satellite quenched fraction is lower than in the simulations across the $M_*$ range considered.\ The difference is largest for $10^7 \lesssim M_*/\msun \lesssim 10^8$, where the \SAGAII{} spectroscopic coverage is essentially complete and the incompleteness/interloper-corrected observed quenched fraction is 2--3 times lower than in the simulations.\ We discuss the implications of this result in the following section.

Finally, the median projected separation for the \SAGAII{} quenched and star-forming objects are shown at the bottom of Figure \ref{fig:satcomp} by short, white lines.\ These separations are relatively consistent with the simulations where star-forming satellites have larger projected separations than quenched satellites, however the scarcity and incompleteness of observed quenched satellites limits this comparison.\ This perspective illustrates a distance dependence in both simulated samples beginning at intermediate masses $(10^8 \lesssim M_*/\msun\lesssim 10^9$) where quenched satellites are located at lower projected distances and become more ubiquitous at lower stellar masses, similar to previous trends reported around more massive hosts and in the LG \citep[][]{2013Guo,2014Wang,2018Fillingham}.

\section{Discussion and Conclusions} \label{sec:Conclusion}

We have identified star-forming satellites around MW analogs in observed and simulated samples, which have similar sizes, similar host masses, and which are selected in a similar manner (\S \ref{sec:sample}).\ We used UV emission in confirmed \SAGAII{} satellites (Figure \ref{fig:cogexample}) and an instantaneous SFR in APOSTLE and Auriga satellites to define observed and simulated star-forming objects, respectively, which were checked for consistency (\S \ref{sec:Obsanddata} and Figure \ref{fig:sfr-stellar}).\ We compared quenched fractions in the resulting samples, and find that the incompleteness/interloper-corrected observed values are $\sim$2--3 times lower than the simulated ones for $10^7\lesssim M_*/\msun \lesssim 10^8$ (\S \ref{sec:results} and Figures \ref{fig:quenchfracsfhist}~and~\ref{fig:satcomp}).\ The observed and simulated quenched fractions are therefore strongly discrepant in a mass range that is well-probed in both samples.\ 

The comparisons presented here are broadly consistent with previous investigations of satellite quenched fractions.\ Observationally, our quenched fractions differ only slightly from those found by \citetalias{SAGAII}, which \citet{Akins2021} report to be lower than those of satellites around 4 MW-like hosts in the Justice League simulations.\ Furthermore, we note that this discrepancy extends to other nearby systems that are similar to the MW/M31 with respect to quenched satellites \citep[][]{2013Chiboucas,2021Carlsten} and it can also be inferred from the star formation histories of satellites from the FIRE-2 simulations \citep[][]{2019GarrisonKimmel}.\ More broadly, since the quenched fractions in the LG have been shown to agree with simulations \citep[e.g.][]{2016Fillingham, 2018Simpson,2021Joshi,Akins2021} and the LG and \SAGAII{} have been shown to disagree (\citealt{SAGAI}; \citetalias{SAGAII}), the discrepancy in quenched fractions between \SAGAII{} and the APOSTLE and Auriga simulations is unsurprising.\ Here, we have demonstrated the degree to which the observations and simulations are inconsistent with large, comparably-sized samples, and that the discrepancy is robust against different choices of observed or simulated star formation tracers as well as a variety of simulation parameters (\S \ref{subsec:simSF} and Appendix~\ref{sec:restests}).

The agreement between the SFR--$M_*$ relations and cumulative $M_*$ distributions in Figure \ref{fig:sfr-stellar} combined with the discrepant quenched fractions in Figures \ref{fig:quenchfracsfhist}~and~\ref{fig:satcomp} suggest that the difference between the observed and simulated samples stems from the number of quenched satellites in each.\ For $10^{6.6} \lesssim M_*/\msun \lesssim 10^{7.8}$, the APOSTLE and Auriga samples, respectively, have an average of $3.5$ and $4.9$ satellites per host, comparable to the LG \citep[][]{2012McConnachie} but greater than the incompleteness/interloper-corrected \SAGAII{} value\footnote{The incompleteness/interloper models of \citetalias{SAGAII} predict that Table~\ref{table:maintable} is missing $\sim0.7$ satellites per host in the $10^{6.6}\lesssim M_*/\msun \lesssim 10^{7.8}$ range; see their \S 5.3.} of $\sim$2 \citepalias{SAGAII}.\ The star-forming satellite counts, quenched fractions and total satellite counts presented here are therefore all broadly consistent at $10^{6.6}\lesssim M_*/\msun \lesssim 10^{7.8}$ if there are $50 - 100$ additional quenched satellites in the simulated samples compared to that estimated for the incompleteness/interloper-corrected observed sample.\

One possibility is that the quenched satellite number difference is observationally-driven: in this scenario, the \SAGAII{} incompleteness correction under-estimates the number of quenched satellites with $10^{6.6} \lesssim M_*/\msun \lesssim 10^{7.8}$, with the deficit increasing towards the low-mass end (c.f.~Figure \ref{fig:sfr-stellar} right).\ The nearly complete \SAGAII{} spectroscopic coverage and the conservative \citetalias{SAGAII} incompleteness correction suggest that the quenched satellites would most likely be missing from the imaging catalogs from which spectroscopic targets are drawn. 

It is plausible that low surface brightness (LSB) satellites are missing from the \SAGAII{} imaging catalogs from which follow-up targets are drawn since they are not developed for LSB detection.\ \citetalias{SAGAII}'s comparisons to deeper overlapping catalogs argue against this scenario, although quantitative photometric completeness simulations \citep[e.g.][]{2017Bennet} have not been carried out.\ It is also plausible that a larger fraction of the 70 detected diffuse LSB galaxies (dLSBGs) without redshifts are actually satellites than the $25\% - 30\%$ assumed in the \citetalias{SAGAII} incompleteness correction.\ If all of these dLSBGs were satellites, it may remedy the discrepancy at the lowest masses, however, the total number of quenched satellites per host would still be low compared to that in the simulations.\ Extending these investigations of surface brightness effects to simulations may provide some additional insight \citep[e.g.][]{2020Font}.\ We conclude that observational effects are unlikely to fully explain the quenched fraction discrepancy reported here.

A second possibility is that the observed and simulated quenched fraction difference is simulation-driven: in this scenario, the simulations over-predict the number of quenched satellites around MW analogs.\ The correspondence between APOSTLE and Auriga in Figures \ref{fig:quenchfracsfhist}~and~\ref{fig:satcomp} as well as similar results from other simulations \citep[][]{Akins2021,2021Joshi} imply that the effect is somewhat model-agnostic.\ This consistency is not necessarily predictable: while tides are relatively similar across simulation suites, the interstellar medium (ISM), star-formation feedback-dependent physics and hydrodynamical schemes producing the ram pressure that begins to dominate quenching of $10^7 \lesssim M_*/\msun \lesssim 10^8$ satellites \citep[][]{2015Wetzel,2016Fillingham} are not \citep[e.g.][]{2007Agertz,2012Sijacki}.\ Furthermore, \citet[][]{2019Digby} find all intermediate-mass ($10^7 \lesssim M_*/\msun \lesssim 10^9$) dwarfs in APOSTLE and Auriga have young ($\tau_{\rm form}\lesssim$ 6 Gyr ago) stellar populations.\ This suggests that any form of quenching in these satellites, as implied in this work, must have occurred recently and rapidly, consistent with previous similar investigations \citep[e.g.][]{2015Wetzel,2016Fillingham}.\

Whether or not the agreement between simulated quenched fractions presented here has a common physical origin or stems from a confluence of disparate effects with a similar net outcome is unclear.\ Nonetheless, it is plausible that the ISM gas densities simulated here with state of the art resolution and star formation feedback physics in the simulations generically produce satellites that are less resilient to ram-pressure stripping than in nature.\ A separate, detailed study is required to determine if this mechanism quantitatively explains the discrepancy reported here \citep[e.g.][]{2019Bose,2019Digby}.

 We conclude that the dearth of observed quenched satellites relative to simulated ones in Figures \ref{fig:quenchfracsfhist}~and~\ref{fig:satcomp} is not readily explained by vagaries in the samples considered here.\ There is apparently a genuine discrepancy between the satellite populations of the MW, M31 and their simulated analogs on the one hand, and of the \SAGAII{} host galaxies on the other.\ This highlights that while the ability to reproduce the properties of the LG is a necessary feature of any complete model of galaxy formation and evolution, exclusive reliance on the LG as the benchmark for faint satellites risks introducing severe biases in the models.\ More detailed comparisons between observed and simulated satellites will further elucidate the origin of this discrepancy.\ This requires larger, more observationally complete samples that probe even further down the luminosity function \citep[e.g][]{2019Bennet,2020Carlsten}, and large samples of simulated satellites \citep[e.g.\ this work;][]{2020Font,2021Joshi} at higher resolutions \citep[e.g.][]{2016Wetzel,2019Wheeler} and self-consistent star-forming ISMs.\ Both will be available soon.

\floattable
\begin{deluxetable}{ccccccCcCCcc}[h!]
\tablecaption{UV properties of Observed satellites\label{table:maintable}}
%\rotate
\tabletypesize{\tiny}
\tablehead{
\colhead{Name} & \colhead{RA} & \colhead{Dec} & \colhead{$D_{host}$} & \colhead{$r_{COG}$} & \colhead{$(S/N)_{NUV}$} & \colhead{$m_{NUV}$} & \colhead{$(S/N)_{FUV}$} & \colhead{$m_{FUV}$} & \colhead{$\sim\mathrm{log}SFR_{NUV}$} & \colhead{SF?} &\colhead{GALEX Tile} \\
\colhead{} & \colhead{deg} & \colhead{deg} & \colhead{(Mpc)} & \colhead{(arcsec)} & \colhead{} & \colhead{(mag)} & \colhead{} & \colhead{(mag)} & \colhead{$(M_{\odot}yr^{-1})$} & \colhead{} & \colhead{} \\
\colhead{(1)} & \colhead{(2)} & \colhead{(3)} & \colhead{(4)} & \colhead{(5)} & \colhead{(6)} & \colhead{(7)} & \colhead{(8)} & \colhead{(9)} & \colhead{(10)} & \colhead{(11)} & \colhead{(12)}
} 
\startdata
LS-429811-3398 & 20.285 & 17.6022 & 38.4 & 6.8 & 5.7 & 20.78\pm0.20 & 8.4 & 20.55\pm0.14 & -2.57 & Y & AIS\textunderscore183\textunderscore50183\textunderscore0001\textunderscore sv27\\
LS-431187-1672 & 20.328 & 17.7539 & 38.4 & 9.3 & 5.7 & 20.49\pm0.19 & 5.7 & 20.61\pm0.19 & -2.46 & Y & AIS\textunderscore183\textunderscore50183\textunderscore0001\textunderscore sv18\\
LS-429812-2469 & 20.5362 & 17.5279 & 38.4 & 7.2 & 7.2 & 20.50\pm0.16 & 6.6 & 20.72\pm0.18 & -2.46 & Y & AIS\textunderscore183\textunderscore50183\textunderscore0001\textunderscore sv27\\
LS-432563-224 & 20.7772 & 17.8916 & 38.4 & 13 & 13.9 & 18.88\pm0.08 & 18.9 & 19.01\pm0.06 & -1.81& Y & AIS\textunderscore183\textunderscore50183\textunderscore0001\textunderscore sv26\\
DES-313240666 & 39.9254 & -1.4187 & 37 & 8.8 & 13.9 & 19.40\pm0.08 & \nodata & \nodata & -2.05 & Y & AIS\textunderscore284\textunderscore50284\textunderscore0001\textunderscore sv49\\
DES-350665706 & 50.1913 & -15.5749 & 34.3 & 10.6 & 23.2 & 19.49\pm0.05 & 22 & 19.73\pm0.05 & -2.16 & Y & MISWZS03\textunderscore27553\textunderscore0283\textunderscore17492\\
DES-353757883 & 50.4652 & -15.7104 & 34.3 & 61.7 & 21.5 & 16.16\pm0.05 & 32.2 & 16.36\pm0.03 & -0.82\ & Y & MISWZS03\textunderscore27605\textunderscore0283\textunderscore17497\\
DES-353742769 & 50.9464 & -15.4004 & 34.3 & 9.2 & 8.6 & 21.58\pm0.13 & 1.2 & >24.04 & -2.99 & Y & MISWZS03\textunderscore27552\textunderscore0283\textunderscore17564\\
DES-371747881 & 55.3397 & -13.1446 & 31.9 & 15.3 & 34.7 & 17.67\pm0.03 & 29.2 & 17.88\pm0.04 & -1.49 & Y & AIS\textunderscore182\textunderscore50182\textunderscore0001\textunderscore sv68\\
DES-373383928 & 55.5682 & -13.217 & 31.9 & 9.1 & 21.6 & 18.35\pm0.05 & 22.1 & 18.55\pm0.05 & -1.76 & Y & AIS\textunderscore182\textunderscore50182\textunderscore0001\textunderscore sv68\\
DES-373393030 & 55.5841 & -13.4218 & 31.9 & 6.8 & 16.2 & 19.48\pm0.07 & 13 & 19.55\pm0.08 & -2.22 & Y & AIS\textunderscore182\textunderscore50182\textunderscore0001\textunderscore sv68 \\
\enddata

\tablecomments{The first 10 rows of this table are shown here.\ The full table is available online in machine readable format.\ \\ \\ cols.~(1)--(4): \citetalias{SAGAII} satellite name, J2000 centroid position and host distance.\ col.~(5) Curve-of-growth radius around the optical centroid used to measure UV fluxes.\ col.~(6): NUV signal-to-noise ratio.\ col.(7): NUV apparent AB magnitude, corrected for foreground extinction.\ col.~(8)-(9): Same as cols.~(6) and(7) but for FUV.\ col.(10) First-order NUV star formation rates computed using fluxes in col.~(7), distances from col.~(4) and Equation (3) from \citet[][]{2006IglesiasParamo} uncorrected for internal dust attenuation.\ col.(11): Star-forming classification as defined in \S \ref{subsec:obsSF} col.(12): name of {\it GALEX} tile used.\ }
\end{deluxetable}

\acknowledgments
We thank the anonymous referee for their thoughtful and useful comments that helped improve this manuscript.\ KS acknowledges support from the Natural Sciences and Engineering Research Council of Canada (NSERC).\ KAO acknowledges support by the European
Research Council (ERC) through Advanced Investigator grant to C.S.\ Frenk, DMIDAS (GA 786910).\ AF is supported by a UKRI Future Leaders Fellowship (grant no MR/T042362/1).\
The simulation used in this work used the DiRAC@Durham facility managed by the Institute for Computational Cosmology on behalf of the STFC DiRAC HPC Facility (\url{www.dirac.ac.uk}).\ The equipment was funded by BEIS capital funding via STFC capital grants ST/K00042X/1, ST/P002293/1, ST/R002371/1 and ST/S002502/1, Durham University and STFC operations grant ST/R000832/1.\ DiRAC is part of the National e-Infrastructure.\ DJS acknowledges support from NSF grants AST-1821967 and 1813708.\ FAG acknowledges financial support from CONICYT through the project FONDECYT Regular Nr.\ 1211370.\ FAG acknowledge funding from the Max Planck Society through a Partner Group grant.\ FM acknowledges support through the Program ``Rita Levi Montalcini" of the Italian MUR.\ B.M.P.\ is supported by an NSF Astronomy and Astrophysics Postdoctoral Fellowship under award AST-2001663.

This research made use of data from the SAGA Survey (sagasurvey.org).\ The SAGA Survey was supported by NSF collaborative grants AST-1517148 and AST-1517422 and by Heising-Simons Foundation grant 2019-1402.

The Legacy Surveys consist of three individual and complementary projects: the Dark Energy Camera Legacy Survey (DECaLS; NOAO Proposal ID \# 2014B-0404; PIs: David Schlegel and Arjun Dey), the Beijing-Arizona Sky Survey (BASS; NOAO Proposal ID \# 2015A-0801; PIs: Zhou Xu and Xiaohui Fan), and the Mayall z-band Legacy Survey (MzLS; NOAO Proposal ID \# 2016A-0453; PI: Arjun Dey).\ DECaLS, BASS and MzLS together include data obtained, respectively, at the Blanco telescope, Cerro Tololo Inter-American Observatory, National Optical Astronomy Observatory (NOAO); the Bok telescope, Steward Observatory, University of Arizona; and the Mayall telescope, Kitt Peak National Observatory, NOAO.\ The Legacy Surveys project is honored to be permitted to conduct astronomical research on Iolkam Du'ag (Kitt Peak), a mountain with particular significance to the Tohono O'odham Nation.

NOAO is operated by the Association of Universities for Research in Astronomy (AURA) under a cooperative agreement with the National Science Foundation.

This project used data obtained with the Dark Energy Camera (DECam), which was constructed by the Dark Energy Survey (DES) collaboration.\ Funding for the DES Projects has been provided by the U.S.\ Department of Energy, the U.S.\ National Science Foundation, the Ministry of Science and Education of Spain, the Science and Technology Facilities Council of the United Kingdom, the Higher Education Funding Council for England, the National Center for Supercomputing Applications at the University of Illinois at Urbana-Champaign, the Kavli Institute of Cosmological Physics at the University of Chicago, Center for Cosmology and Astro-Particle Physics at the Ohio State University, the Mitchell Institute for Fundamental Physics and Astronomy at Texas A\&M University, Financiadora de Estudos e Projetos, Fundacao Carlos Chagas Filho de Amparo, Financiadora de Estudos e Projetos, Fundacao Carlos Chagas Filho de Amparo a Pesquisa do Estado do Rio de Janeiro, Conselho Nacional de Desenvolvimento Cientifico e Tecnologico and the Ministerio da Ciencia, Tecnologia e Inovacao, the Deutsche Forschungsgemeinschaft and the Collaborating Institutions in the Dark Energy Survey.\ The Collaborating Institutions are Argonne National Laboratory, the University of California at Santa Cruz, the University of Cambridge, Centro de Investigaciones Energeticas, Medioambientales y Tecnologicas-Madrid, the University of Chicago, University College London, the DES-Brazil Consortium, the University of Edinburgh, the Eidgenossische Technische Hochschule (ETH) Zurich, Fermi National Accelerator Laboratory, the University of Illinois at Urbana-Champaign, the Institut de Ciencies de l'Espai (IEEC/CSIC), the Institut de Fisica d'Altes Energies, Lawrence Berkeley National Laboratory, the Ludwig-Maximilians Universitat Munchen and the associated Excellence Cluster Universe, the University of Michigan, the National Optical Astronomy Observatory, the University of Nottingham, the Ohio State University, the University of Pennsylvania, the University of Portsmouth, SLAC National Accelerator Laboratory, Stanford University, the University of Sussex, and Texas A\&M University.

BASS is a key project of the Telescope Access Program (TAP), which has been funded by the National Astronomical Observatories of China, the Chinese Academy of Sciences (the Strategic Priority Research Program "The Emergence of Cosmological Structures" Grant \# XDB09000000), and the Special Fund for Astronomy from the Ministry of Finance.\ The BASS is also supported by the External Cooperation Program of Chinese Academy of Sciences (Grant \# 114A11KYSB20160057), and Chinese National Natural Science Foundation (Grant \# 11433005).

The Legacy Survey team makes use of data products from the Near-Earth Object Wide-field Infrared Survey Explorer (NEOWISE), which is a project of the Jet Propulsion Laboratory/California Institute of Technology.\ NEOWISE is funded by the National Aeronautics and Space Administration.

The Legacy Surveys imaging of the DESI footprint is supported by the Director, Office of Science, Office of High Energy Physics of the U.S.\ Department of Energy under Contract No.\ DE-AC02-05CH1123, by the National Energy Research Scientific Computing Center, a DOE Office of Science User Facility under the same contract; and by the U.S.\ National Science Foundation, Division of Astronomical Sciences under Contract No.\ AST-0950945 to NOAO.

\vspace{5mm}

\software{astropy \citep{astropy:2013,astropy:2018}; Photutils \citep[][]{larrybradley2020}}
\facility{{\it GALEX}, DiRAC}

\bibliographystyle{aasjournal}
\bibliography{references}

\begin{thebibliography}{}
\expandafter\ifx\csname natexlab\endcsname\relax\def\natexlab#1{#1}\fi
\providecommand{\url}[1]{\href{#1}{#1}}
\providecommand{\dodoi}[1]{doi:~\href{http://doi.org/#1}{\nolinkurl{#1}}}
\providecommand{\doeprint}[1]{\href{http://ascl.net/#1}{\nolinkurl{http://ascl.net/#1}}}
\providecommand{\doarXiv}[1]{\href{https://arxiv.org/abs/#1}{\nolinkurl{https://arxiv.org/abs/#1}}}

\bibitem[{{Agertz} {et~al.}(2007){Agertz}, {Moore}, {Stadel}, {Potter},
  {Miniati}, {Read}, {Mayer}, {Gawryszczak}, {Kravtsov}, {Nordlund}, {Pearce},
  {Quilis}, {Rudd}, {Springel}, {Stone}, {Tasker}, {Teyssier}, {Wadsley}, \&
  {Walder}}]{2007Agertz}
{Agertz}, O., {Moore}, B., {Stadel}, J., {et~al.} 2007, \mnras, 380, 963,
  \dodoi{10.1111/j.1365-2966.2007.12183.x}

\bibitem[{{Akins} {et~al.}(2021){Akins}, {Christensen}, {Brooks}, {Munshi},
  {Applebaum}, {Engelhardt}, \& {Chamberland}}]{Akins2021}
{Akins}, H.~B., {Christensen}, C.~R., {Brooks}, A.~M., {et~al.} 2021, \apj,
  909, 139, \dodoi{10.3847/1538-4357/abe2ab}

\bibitem[{{Astropy Collaboration} {et~al.}(2013){Astropy Collaboration},
  {Robitaille}, {Tollerud}, {Greenfield}, {Droettboom}, {Bray}, {Aldcroft},
  {Davis}, {Ginsburg}, {Price-Whelan}, {Kerzendorf}, {Conley}, {Crighton},
  {Barbary}, {Muna}, {Ferguson}, {Grollier}, {Parikh}, {Nair}, {Unther},
  {Deil}, {Woillez}, {Conseil}, {Kramer}, {Turner}, {Singer}, {Fox}, {Weaver},
  {Zabalza}, {Edwards}, {Azalee Bostroem}, {Burke}, {Casey}, {Crawford},
  {Dencheva}, {Ely}, {Jenness}, {Labrie}, {Lim}, {Pierfederici}, {Pontzen},
  {Ptak}, {Refsdal}, {Servillat}, \& {Streicher}}]{astropy:2013}
{Astropy Collaboration}, {Robitaille}, T.~P., {Tollerud}, E.~J., {et~al.} 2013,
  \aap, 558, A33, \dodoi{10.1051/0004-6361/201322068}

\bibitem[{{Bennet} {et~al.}(2019){Bennet}, {Sand}, {Crnojevi{\'c}}, {Spekkens},
  {Karunakaran}, {Zaritsky}, \& {Mutlu-Pakdil}}]{2019Bennet}
{Bennet}, P., {Sand}, D.~J., {Crnojevi{\'c}}, D., {et~al.} 2019, \apj, 885,
  153, \dodoi{10.3847/1538-4357/ab46ab}

\bibitem[{{Bennet} {et~al.}(2017){Bennet}, {Sand}, {Crnojevi{\'c}}, {Spekkens},
  {Zaritsky}, \& {Karunakaran}}]{2017Bennet}
---. 2017, \apj, 850, 109, \dodoi{10.3847/1538-4357/aa9180}

\bibitem[{{Bose} {et~al.}(2019){Bose}, {Frenk}, {Jenkins}, {Fattahi},
  {G{\'o}mez}, {Grand}, {Marinacci}, {Navarro}, {Oman}, {Pakmor}, {Schaye},
  {Simpson}, \& {Springel}}]{2019Bose}
{Bose}, S., {Frenk}, C.~S., {Jenkins}, A., {et~al.} 2019, \mnras, 486, 4790,
  \dodoi{10.1093/mnras/stz1168}

\bibitem[{{Bradley} {et~al.}(2020){Bradley}, {Sip\~ocz}, {Robitaille},
  {Tollerud}, {Vinícius}, {Deil}, {Barbary}, {Wilson}, {Busko}, {G\"unther},
  {Cara}, {Conseil}, {Bostroem}, {Droettboom}, {Bray}, {Bratholm}, {Lim},
  {Barentsen}, {Craig}, {Pascual}, {Perren}, {Greco}, {Donath}, {Val-Borro},
  {Kerzendorf}, {Bach}, {Weaver}, {D'Eugenio}, {Souchereau}, \&
  {Ferreira}}]{larrybradley2020}
{Bradley}, L., {Sip\~ocz}, B., {Robitaille}, T., {et~al.} 2020,
  astropy/photutils: 1.0.1, 1.0.1,  Zenodo, \dodoi{10.5281/zenodo.4049061}.
\newblock \url{https://doi.org/10.5281/zenodo.4049061}

\bibitem[{Brown {et~al.}(2001)Brown, Cai, \& DasGupta}]{brown2001interval}
Brown, L.~D., Cai, T.~T., \& DasGupta, A. 2001, Statistical science, 101,
  \dodoi{10.1214/ss/1009213286}

\bibitem[{{Carlsten} {et~al.}(2020){Carlsten}, {Greco}, {Beaton}, \&
  {Greene}}]{2020Carlsten}
{Carlsten}, S.~G., {Greco}, J.~P., {Beaton}, R.~L., \& {Greene}, J.~E. 2020,
  \apj, 891, 144, \dodoi{10.3847/1538-4357/ab7758}

\bibitem[{{Carlsten} {et~al.}(2021){Carlsten}, {Greene}, {Greco}, {Beaton}, \&
  {Kado-Fong}}]{2021Carlsten}
{Carlsten}, S.~G., {Greene}, J.~E., {Greco}, J.~P., {Beaton}, R.~L., \&
  {Kado-Fong}, E. 2021, arXiv e-prints, arXiv:2105.03435.
\newblock \doarXiv{2105.03435}

\bibitem[{{Chiboucas} {et~al.}(2013){Chiboucas}, {Jacobs}, {Tully}, \&
  {Karachentsev}}]{2013Chiboucas}
{Chiboucas}, K., {Jacobs}, B.~A., {Tully}, R.~B., \& {Karachentsev}, I.~D.
  2013, \aj, 146, 126, \dodoi{10.1088/0004-6256/146/5/126}

\bibitem[{{Crain} {et~al.}(2015){Crain}, {Schaye}, {Bower}, {Furlong},
  {Schaller}, {Theuns}, {Dalla Vecchia}, {Frenk}, {McCarthy}, {Helly},
  {Jenkins}, {Rosas-Guevara}, {White}, \& {Trayford}}]{2015Crain}
{Crain}, R.~A., {Schaye}, J., {Bower}, R.~G., {et~al.} 2015, \mnras, 450, 1937,
  \dodoi{10.1093/mnras/stv725}

\bibitem[{{Crnojevi{\'c}} {et~al.}(2016){Crnojevi{\'c}}, {Sand}, {Spekkens},
  {Caldwell}, {Guhathakurta}, {McLeod}, {Seth}, {Simon}, {Strader}, \&
  {Toloba}}]{2016Crnojevic}
{Crnojevi{\'c}}, D., {Sand}, D.~J., {Spekkens}, K., {et~al.} 2016, \apj, 823,
  19, \dodoi{10.3847/0004-637X/823/1/19}

\bibitem[{{Davies} {et~al.}(2019){Davies}, {Robotham}, {Lagos}, {Driver},
  {Stevens}, {Bah{\'e}}, {Alpaslan}, {Bremer}, {Brown}, {Brough},
  {Bland-Hawthorn}, {Cortese}, {Elahi}, {Grootes}, {Holwerda}, {Ludlow},
  {McGee}, {Owers}, \& {Phillipps}}]{2019Davies}
{Davies}, L.~J.~M., {Robotham}, A.~S.~G., {Lagos}, C. d.~P., {et~al.} 2019,
  \mnras, 483, 5444, \dodoi{10.1093/mnras/sty3393}

\bibitem[{{Dey} {et~al.}(2019){Dey}, {Schlegel}, {Lang}, {Blum}, {Burleigh},
  {Fan}, {Findlay}, {Finkbeiner}, {Herrera}, {Juneau}, {Landriau}, {Levi},
  {McGreer}, {Meisner}, {Myers}, {Moustakas}, {Nugent}, {Patej}, {Schlafly},
  {Walker}, {Valdes}, {Weaver}, {Y{\`e}che}, {Zou}, {Zhou}, {Abareshi},
  {Abbott}, {Abolfathi}, {Aguilera}, {Alam}, {Allen}, {Alvarez}, {Annis},
  {Ansarinejad}, {Aubert}, {Beechert}, {Bell}, {BenZvi}, {Beutler}, {Bielby},
  {Bolton}, {Brice{\~n}o}, {Buckley-Geer}, {Butler}, {Calamida}, {Carlberg},
  {Carter}, {Casas}, {Castander}, {Choi}, {Comparat}, {Cukanovaite}, {Delubac},
  {DeVries}, {Dey}, {Dhungana}, {Dickinson}, {Ding}, {Donaldson}, {Duan},
  {Duckworth}, {Eftekharzadeh}, {Eisenstein}, {Etourneau}, {Fagrelius},
  {Farihi}, {Fitzpatrick}, {Font-Ribera}, {Fulmer}, {G{\"a}nsicke},
  {Gaztanaga}, {George}, {Gerdes}, {Gontcho}, {Gorgoni}, {Green}, {Guy},
  {Harmer}, {Hernandez}, {Honscheid}, {Huang}, {James}, {Jannuzi}, {Jiang},
  {Joyce}, {Karcher}, {Karkar}, {Kehoe}, {Kneib}, {Kueter-Young}, {Lan},
  {Lauer}, {Le Guillou}, {Le Van Suu}, {Lee}, {Lesser}, {Perreault Levasseur},
  {Li}, {Mann}, {Marshall}, {Mart{\'\i}nez-V{\'a}zquez}, {Martini}, {du Mas des
  Bourboux}, {McManus}, {Meier}, {M{\'e}nard}, {Metcalfe},
  {Mu{\~n}oz-Guti{\'e}rrez}, {Najita}, {Napier}, {Narayan}, {Newman}, {Nie},
  {Nord}, {Norman}, {Olsen}, {Paat}, {Palanque-Delabrouille}, {Peng},
  {Poppett}, {Poremba}, {Prakash}, {Rabinowitz}, {Raichoor}, {Rezaie},
  {Robertson}, {Roe}, {Ross}, {Ross}, {Rudnick}, {Safonova}, {Saha},
  {S{\'a}nchez}, {Savary}, {Schweiker}, {Scott}, {Seo}, {Shan}, {Silva},
  {Slepian}, {Soto}, {Sprayberry}, {Staten}, {Stillman}, {Stupak}, {Summers},
  {Sien Tie}, {Tirado}, {Vargas-Maga{\~n}a}, {Vivas}, {Wechsler}, {Williams},
  {Yang}, {Yang}, {Yapici}, {Zaritsky}, {Zenteno}, {Zhang}, {Zhang}, {Zhou}, \&
  {Zhou}}]{2019Dey}
{Dey}, A., {Schlegel}, D.~J., {Lang}, D., {et~al.} 2019, \aj, 157, 168,
  \dodoi{10.3847/1538-3881/ab089d}

\bibitem[{{Digby} {et~al.}(2019){Digby}, {Navarro}, {Fattahi}, {Simpson},
  {Oman}, {Gomez}, {Frenk}, {Grand}, \& {Pakmor}}]{2019Digby}
{Digby}, R., {Navarro}, J.~F., {Fattahi}, A., {et~al.} 2019, \mnras, 485, 5423,
  \dodoi{10.1093/mnras/stz745}

\bibitem[{{Fattahi} {et~al.}(2016){Fattahi}, {Navarro}, {Sawala}, {Frenk},
  {Oman}, {Crain}, {Furlong}, {Schaller}, {Schaye}, {Theuns}, \&
  {Jenkins}}]{2016Fattahi}
{Fattahi}, A., {Navarro}, J.~F., {Sawala}, T., {et~al.} 2016, \mnras, 457, 844,
  \dodoi{10.1093/mnras/stv2970}

\bibitem[{{Fillingham} {et~al.}(2018){Fillingham}, {Cooper}, {Boylan-Kolchin},
  {Bullock}, {Garrison-Kimmel}, \& {Wheeler}}]{2018Fillingham}
{Fillingham}, S.~P., {Cooper}, M.~C., {Boylan-Kolchin}, M., {et~al.} 2018,
  \mnras, 477, 4491, \dodoi{10.1093/mnras/sty958}

\bibitem[{{Fillingham} {et~al.}(2016){Fillingham}, {Cooper}, {Pace},
  {Boylan-Kolchin}, {Bullock}, {Garrison-Kimmel}, \&
  {Wheeler}}]{2016Fillingham}
{Fillingham}, S.~P., {Cooper}, M.~C., {Pace}, A.~B., {et~al.} 2016, \mnras,
  463, 1916, \dodoi{10.1093/mnras/stw2131}

\bibitem[{{Fillingham} {et~al.}(2015){Fillingham}, {Cooper}, {Wheeler},
  {Garrison-Kimmel}, {Boylan-Kolchin}, \& {Bullock}}]{2015fillingham}
{Fillingham}, S.~P., {Cooper}, M.~C., {Wheeler}, C., {et~al.} 2015, \mnras,
  454, 2039, \dodoi{10.1093/mnras/stv2058}

\bibitem[{{Font} {et~al.}(2020){Font}, {McCarthy}, \& {Belokurov}}]{2020Font}
{Font}, A.~S., {McCarthy}, I.~G., \& {Belokurov}, V. 2020, arXiv e-prints,
  arXiv:2011.12974.
\newblock \doarXiv{2011.12974}

\bibitem[{{Furlong} {et~al.}(2015){Furlong}, {Bower}, {Theuns}, {Schaye},
  {Crain}, {Schaller}, {Dalla Vecchia}, {Frenk}, {McCarthy}, {Helly},
  {Jenkins}, \& {Rosas-Guevara}}]{2015Furlong}
{Furlong}, M., {Bower}, R.~G., {Theuns}, T., {et~al.} 2015, \mnras, 450, 4486,
  \dodoi{10.1093/mnras/stv852}

\bibitem[{{Garrison-Kimmel} {et~al.}(2019){Garrison-Kimmel}, {Wetzel},
  {Hopkins}, {Sanderson}, {El-Badry}, {Graus}, {Chan}, {Feldmann},
  {Boylan-Kolchin}, {Hayward}, {Bullock}, {Fitts}, {Samuel}, {Wheeler},
  {Kere{\v{s}}}, \& {Faucher-Gigu{\`e}re}}]{2019GarrisonKimmel}
{Garrison-Kimmel}, S., {Wetzel}, A., {Hopkins}, P.~F., {et~al.} 2019, \mnras,
  489, 4574, \dodoi{10.1093/mnras/stz2507}

\bibitem[{{Geha} {et~al.}(2012){Geha}, {Blanton}, {Yan}, \&
  {Tinker}}]{2012Geha}
{Geha}, M., {Blanton}, M.~R., {Yan}, R., \& {Tinker}, J.~L. 2012, \apj, 757,
  85, \dodoi{10.1088/0004-637X/757/1/85}

\bibitem[{{Geha} {et~al.}(2017){Geha}, {Wechsler}, {Mao}, {Tollerud}, {Weiner},
  {Bernstein}, {Hoyle}, {Marchi}, {Marshall}, {Mu{\~n}oz}, \& {Lu}}]{SAGAI}
{Geha}, M., {Wechsler}, R.~H., {Mao}, Y.-Y., {et~al.} 2017, \apj, 847, 4,
  \dodoi{10.3847/1538-4357/aa8626}

\bibitem[{{Grand} {et~al.}(2017){Grand}, {G{\'o}mez}, {Marinacci}, {Pakmor},
  {Springel}, {Campbell}, {Frenk}, {Jenkins}, \& {White}}]{2017Grand}
{Grand}, R. J.~J., {G{\'o}mez}, F.~A., {Marinacci}, F., {et~al.} 2017, \mnras,
  467, 179, \dodoi{10.1093/mnras/stx071}

\bibitem[{{Grand} {et~al.}(2019){Grand}, {van de Voort}, {Zjupa}, {Fragkoudi},
  {G{\'o}mez}, {Kauffmann}, {Marinacci}, {Pakmor}, {Springel}, \&
  {White}}]{2019Grand}
{Grand}, R. J.~J., {van de Voort}, F., {Zjupa}, J., {et~al.} 2019, \mnras, 490,
  4786, \dodoi{10.1093/mnras/stz2928}

\bibitem[{{Grcevich} \& {Putman}(2009)}]{HILocalGroup}
{Grcevich}, J., \& {Putman}, M.~E. 2009, \apj, 696, 385,
  \dodoi{10.1088/0004-637X/696/1/385}

\bibitem[{{Guo} {et~al.}(2013){Guo}, {Cole}, {Eke}, {Frenk}, \&
  {Helly}}]{2013Guo}
{Guo}, Q., {Cole}, S., {Eke}, V., {Frenk}, C., \& {Helly}, J. 2013, \mnras,
  434, 1838, \dodoi{10.1093/mnras/stt903}

\bibitem[{{Iglesias-P{\'a}ramo} {et~al.}(2006){Iglesias-P{\'a}ramo}, {Buat},
  {Takeuchi}, {Xu}, {Boissier}, {Boselli}, {Burgarella}, {Madore}, {Gil de
  Paz}, {Bianchi}, {Barlow}, {Byun}, {Donas}, {Forster}, {Friedman}, {Heckman},
  {Jelinski}, {Lee}, {Malina}, {Martin}, {Milliard}, {Morrissey}, {Neff},
  {Rich}, {Schiminovich}, {Seibert}, {Siegmund}, {Small}, {Szalay}, {Welsh}, \&
  {Wyder}}]{2006IglesiasParamo}
{Iglesias-P{\'a}ramo}, J., {Buat}, V., {Takeuchi}, T.~T., {et~al.} 2006, \apjs,
  164, 38, \dodoi{10.1086/502628}

\bibitem[{{Javanmardi} {et~al.}(2016){Javanmardi}, {Martinez-Delgado},
  {Kroupa}, {Henkel}, {Crawford}, {Teuwen}, {Gabany}, {Hanson}, {Chonis}, \&
  {Neyer}}]{2016Javanmardi}
{Javanmardi}, B., {Martinez-Delgado}, D., {Kroupa}, P., {et~al.} 2016, \aap,
  588, A89, \dodoi{10.1051/0004-6361/201527745}

\bibitem[{{Joshi} {et~al.}(2021){Joshi}, {Pillepich}, {Nelson}, {Zinger},
  {Marinacci}, {Springel}, {Vogelsberger}, \& {Hernquist}}]{2021Joshi}
{Joshi}, G.~D., {Pillepich}, A., {Nelson}, D., {et~al.} 2021, arXiv e-prints,
  arXiv:2101.12226.
\newblock \doarXiv{2101.12226}

\bibitem[{{Lee} {et~al.}(2011){Lee}, {Gil de Paz}, {Kennicutt}, {Bothwell},
  {Dalcanton}, {Jos{\'e} G. Funes S.}, {Johnson}, {Sakai}, {Skillman},
  {Tremonti}, \& {van Zee}}]{2011Lee}
{Lee}, J.~C., {Gil de Paz}, A., {Kennicutt}, Robert~C., J., {et~al.} 2011,
  \apjs, 192, 6, \dodoi{10.1088/0067-0049/192/1/6}

\bibitem[{{Mao} {et~al.}(2021){Mao}, {Geha}, {Wechsler}, {Weiner}, {Tollerud},
  {Nadler}, \& {Kallivayalil}}]{SAGAII}
{Mao}, Y.-Y., {Geha}, M., {Wechsler}, R.~H., {et~al.} 2021, \apj, 907, 85,
  \dodoi{10.3847/1538-4357/abce58}

\bibitem[{{Martin} {et~al.}(2005){Martin}, {Fanson}, {Schiminovich},
  {Morrissey}, {Friedman}, {Barlow}, {Conrow}, {Grange}, {Jelinsky},
  {Milliard}, {Siegmund}, {Bianchi}, {Byun}, {Donas}, {Forster}, {Heckman},
  {Lee}, {Madore}, {Malina}, {Neff}, {Rich}, {Small}, {Surber}, {Szalay},
  {Welsh}, \& {Wyder}}]{2005MartinGALEX}
{Martin}, D.~C., {Fanson}, J., {Schiminovich}, D., {et~al.} 2005, \apjl, 619,
  L1, \dodoi{10.1086/426387}

\bibitem[{{McConnachie}(2012)}]{2012McConnachie}
{McConnachie}, A.~W. 2012, \aj, 144, 4, \dodoi{10.1088/0004-6256/144/1/4}

\bibitem[{{McQuinn} {et~al.}(2015){McQuinn}, {Skillman}, {Dolphin}, \&
  {Mitchell}}]{2015McQuinn}
{McQuinn}, K. B.~W., {Skillman}, E.~D., {Dolphin}, A.~E., \& {Mitchell}, N.~P.
  2015, \apj, 808, 109, \dodoi{10.1088/0004-637X/808/2/109}

\bibitem[{{Morrissey} {et~al.}(2007){Morrissey}, {Conrow}, {Barlow}, {Small},
  {Seibert}, {Wyder}, {Budav{\'a}ri}, {Arnouts}, {Friedman}, {Forster},
  {Martin}, {Neff}, {Schiminovich}, {Bianchi}, {Donas}, {Heckman}, {Lee},
  {Madore}, {Milliard}, {Rich}, {Szalay}, {Welsh}, \&
  {Yi}}]{2007MorrisseyGALEX}
{Morrissey}, P., {Conrow}, T., {Barlow}, T.~A., {et~al.} 2007, \apjs, 173, 682,
  \dodoi{10.1086/520512}

\bibitem[{{Nadler} {et~al.}(2020){Nadler}, {Wechsler}, {Bechtol}, {Mao},
  {Green}, {Drlica-Wagner}, {McNanna}, {Mau}, {Pace}, {Simon}, {Kravtsov},
  {Dodelson}, {Li}, {Riley}, {Wang}, {Abbott}, {Aguena}, {Allam}, {Annis},
  {Avila}, {Bernstein}, {Bertin}, {Brooks}, {Burke}, {Rosell}, {Kind},
  {Carretero}, {Costanzi}, {da Costa}, {De Vicente}, {Desai}, {Evrard},
  {Flaugher}, {Fosalba}, {Frieman}, {Garc{\'\i}a-Bellido}, {Gaztanaga},
  {Gerdes}, {Gruen}, {Gschwend}, {Gutierrez}, {Hartley}, {Hinton}, {Honscheid},
  {Krause}, {Kuehn}, {Kuropatkin}, {Lahav}, {Maia}, {Marshall}, {Menanteau},
  {Miquel}, {Palmese}, {Paz-Chinch{\'o}n}, {Plazas}, {Romer}, {Sanchez},
  {Santiago}, {Scarpine}, {Serrano}, {Smith}, {Soares-Santos}, {Suchyta},
  {Tarle}, {Thomas}, {Varga}, {Walker}, \& {DES Collaboration}}]{2020Nadler}
{Nadler}, E.~O., {Wechsler}, R.~H., {Bechtol}, K., {et~al.} 2020, \apj, 893,
  48, \dodoi{10.3847/1538-4357/ab846a}

\bibitem[{{Phillips} {et~al.}(2014){Phillips}, {Wheeler}, {Boylan-Kolchin},
  {Bullock}, {Cooper}, \& {Tollerud}}]{2014Phillips}
{Phillips}, J.~I., {Wheeler}, C., {Boylan-Kolchin}, M., {et~al.} 2014, \mnras,
  437, 1930, \dodoi{10.1093/mnras/stt2023}

\bibitem[{{Phillips} {et~al.}(2015){Phillips}, {Wheeler}, {Cooper},
  {Boylan-Kolchin}, {Bullock}, \& {Tollerud}}]{2015Phillips}
{Phillips}, J.~I., {Wheeler}, C., {Cooper}, M.~C., {et~al.} 2015, \mnras, 447,
  698, \dodoi{10.1093/mnras/stu2192}

\bibitem[{{Price-Whelan} {et~al.}(2018){Price-Whelan}, {Sip{\H{o}}cz},
  {G{\"u}nther}, {Lim}, {Crawford}, {Conseil}, {Shupe}, {Craig}, {Dencheva},
  {Ginsburg}, {VanderPlas}, {Bradley}, {P{\'e}rez-Su{\'a}rez}, {de Val-Borro},
  {Paper Contributors}, {Aldcroft}, {Cruz}, {Robitaille}, {Tollerud},
  {Coordination Committee}, {Ardelean}, {Babej}, {Bach}, {Bachetti}, {Bakanov},
  {Bamford}, {Barentsen}, {Barmby}, {Baumbach}, {Berry}, {Biscani}, {Boquien},
  {Bostroem}, {Bouma}, {Brammer}, {Bray}, {Breytenbach}, {Buddelmeijer},
  {Burke}, {Calderone}, {Cano Rodr{\'\i}guez}, {Cara}, {Cardoso}, {Cheedella},
  {Copin}, {Corrales}, {Crichton}, {D{\textquoteright}Avella}, {Deil},
  {Depagne}, {Dietrich}, {Donath}, {Droettboom}, {Earl}, {Erben}, {Fabbro},
  {Ferreira}, {Finethy}, {Fox}, {Garrison}, {Gibbons}, {Goldstein}, {Gommers},
  {Greco}, {Greenfield}, {Groener}, {Grollier}, {Hagen}, {Hirst}, {Homeier},
  {Horton}, {Hosseinzadeh}, {Hu}, {Hunkeler}, {Ivezi{\'c}}, {Jain}, {Jenness},
  {Kanarek}, {Kendrew}, {Kern}, {Kerzendorf}, {Khvalko}, {King}, {Kirkby},
  {Kulkarni}, {Kumar}, {Lee}, {Lenz}, {Littlefair}, {Ma}, {Macleod},
  {Mastropietro}, {McCully}, {Montagnac}, {Morris}, {Mueller}, {Mumford},
  {Muna}, {Murphy}, {Nelson}, {Nguyen}, {Ninan}, {N{\"o}the}, {Ogaz}, {Oh},
  {Parejko}, {Parley}, {Pascual}, {Patil}, {Patil}, {Plunkett}, {Prochaska},
  {Rastogi}, {Reddy Janga}, {Sabater}, {Sakurikar}, {Seifert}, {Sherbert},
  {Sherwood-Taylor}, {Shih}, {Sick}, {Silbiger}, {Singanamalla}, {Singer},
  {Sladen}, {Sooley}, {Sornarajah}, {Streicher}, {Teuben}, {Thomas},
  {Tremblay}, {Turner}, {Terr{\'o}n}, {van Kerkwijk}, {de la Vega}, {Watkins},
  {Weaver}, {Whitmore}, {Woillez}, {Zabalza}, \& {Contributors}}]{astropy:2018}
{Price-Whelan}, A.~M., {Sip{\H{o}}cz}, B.~M., {G{\"u}nther}, H.~M., {et~al.}
  2018, \aj, 156, 123, \dodoi{10.3847/1538-3881/aabc4f}

\bibitem[{{Putman} {et~al.}(2021){Putman}, {Zheng}, {Price-Whelan}, {Grcevich},
  {Johnson}, {Tollerud}, \& {Peek}}]{2021Putman}
{Putman}, M.~E., {Zheng}, Y., {Price-Whelan}, A.~M., {et~al.} 2021, arXiv
  e-prints, arXiv:2101.07809.
\newblock \doarXiv{2101.07809}

\bibitem[{{Sawala} {et~al.}(2016){Sawala}, {Frenk}, {Fattahi}, {Navarro},
  {Bower}, {Crain}, {Dalla Vecchia}, {Furlong}, {Helly}, {Jenkins}, {Oman},
  {Schaller}, {Schaye}, {Theuns}, {Trayford}, \& {White}}]{2016Sawala}
{Sawala}, T., {Frenk}, C.~S., {Fattahi}, A., {et~al.} 2016, \mnras, 457, 1931,
  \dodoi{10.1093/mnras/stw145}

\bibitem[{{Schaye} {et~al.}(2015){Schaye}, {Crain}, {Bower}, {Furlong},
  {Schaller}, {Theuns}, {Dalla Vecchia}, {Frenk}, {McCarthy}, {Helly},
  {Jenkins}, {Rosas-Guevara}, {White}, {Baes}, {Booth}, {Camps}, {Navarro},
  {Qu}, {Rahmati}, {Sawala}, {Thomas}, \& {Trayford}}]{2015Schaye}
{Schaye}, J., {Crain}, R.~A., {Bower}, R.~G., {et~al.} 2015, \mnras, 446, 521,
  \dodoi{10.1093/mnras/stu2058}

\bibitem[{{Schlafly} \& {Finkbeiner}(2011)}]{2011Schlaflydust}
{Schlafly}, E.~F., \& {Finkbeiner}, D.~P. 2011, \apj, 737, 103,
  \dodoi{10.1088/0004-637X/737/2/103}

\bibitem[{{Sijacki} {et~al.}(2012){Sijacki}, {Vogelsberger}, {Kere{\v{s}}},
  {Springel}, \& {Hernquist}}]{2012Sijacki}
{Sijacki}, D., {Vogelsberger}, M., {Kere{\v{s}}}, D., {Springel}, V., \&
  {Hernquist}, L. 2012, \mnras, 424, 2999,
  \dodoi{10.1111/j.1365-2966.2012.21466.x}

\bibitem[{{Simpson} {et~al.}(2018){Simpson}, {Grand}, {G{\'o}mez}, {Marinacci},
  {Pakmor}, {Springel}, {Campbell}, \& {Frenk}}]{2018Simpson}
{Simpson}, C.~M., {Grand}, R. J.~J., {G{\'o}mez}, F.~A., {et~al.} 2018, \mnras,
  478, 548, \dodoi{10.1093/mnras/sty774}

\bibitem[{{Slater} \& {Bell}(2014)}]{2014Slater}
{Slater}, C.~T., \& {Bell}, E.~F. 2014, \apj, 792, 141,
  \dodoi{10.1088/0004-637X/792/2/141}

\bibitem[{{Spekkens} {et~al.}(2014){Spekkens}, {Urbancic}, {Mason}, {Willman},
  \& {Aguirre}}]{Spekkens2014}
{Spekkens}, K., {Urbancic}, N., {Mason}, B.~S., {Willman}, B., \& {Aguirre},
  J.~E. 2014, \apjl, 795, L5, \dodoi{10.1088/2041-8205/795/1/L5}

\bibitem[{{Springel} \& {Hernquist}(2003)}]{2003SpringelHernquist}
{Springel}, V., \& {Hernquist}, L. 2003, \mnras, 339, 289,
  \dodoi{10.1046/j.1365-8711.2003.06206.x}

\bibitem[{{Springel} {et~al.}(2001){Springel}, {White}, {Tormen}, \&
  {Kauffmann}}]{2001Springel}
{Springel}, V., {White}, S. D.~M., {Tormen}, G., \& {Kauffmann}, G. 2001,
  \mnras, 328, 726, \dodoi{10.1046/j.1365-8711.2001.04912.x}

\bibitem[{{Vogelsberger} {et~al.}(2013){Vogelsberger}, {Genel}, {Sijacki},
  {Torrey}, {Springel}, \& {Hernquist}}]{2013Vogelsberger}
{Vogelsberger}, M., {Genel}, S., {Sijacki}, D., {et~al.} 2013, \mnras, 436,
  3031, \dodoi{10.1093/mnras/stt1789}

\bibitem[{{Wang} {et~al.}(2014){Wang}, {Sales}, {Henriques}, \&
  {White}}]{2014Wang}
{Wang}, W., {Sales}, L.~V., {Henriques}, B. M.~B., \& {White}, S. D.~M. 2014,
  \mnras, 442, 1363, \dodoi{10.1093/mnras/stu988}

\bibitem[{{Wetzel} {et~al.}(2016){Wetzel}, {Hopkins}, {Kim},
  {Faucher-Gigu{\`e}re}, {Kere{\v{s}}}, \& {Quataert}}]{2016Wetzel}
{Wetzel}, A.~R., {Hopkins}, P.~F., {Kim}, J.-h., {et~al.} 2016, \apjl, 827,
  L23, \dodoi{10.3847/2041-8205/827/2/L23}

\bibitem[{{Wetzel} {et~al.}(2015){Wetzel}, {Tollerud}, \& {Weisz}}]{2015Wetzel}
{Wetzel}, A.~R., {Tollerud}, E.~J., \& {Weisz}, D.~R. 2015, \apjl, 808, L27,
  \dodoi{10.1088/2041-8205/808/1/L27}

\bibitem[{{Wheeler} {et~al.}(2014){Wheeler}, {Phillips}, {Cooper},
  {Boylan-Kolchin}, \& {Bullock}}]{2014Wheeler}
{Wheeler}, C., {Phillips}, J.~I., {Cooper}, M.~C., {Boylan-Kolchin}, M., \&
  {Bullock}, J.~S. 2014, \mnras, 442, 1396, \dodoi{10.1093/mnras/stu965}

\bibitem[{{Wheeler} {et~al.}(2019){Wheeler}, {Hopkins}, {Pace},
  {Garrison-Kimmel}, {Boylan-Kolchin}, {Wetzel}, {Bullock}, {Kere{\v{s}}},
  {Faucher-Gigu{\`e}re}, \& {Quataert}}]{2019Wheeler}
{Wheeler}, C., {Hopkins}, P.~F., {Pace}, A.~B., {et~al.} 2019, \mnras, 490,
  4447, \dodoi{10.1093/mnras/stz2887}

\bibitem[{{Wyder} {et~al.}(2007){Wyder}, {Martin}, {Schiminovich}, {Seibert},
  {Budav{\'a}ri}, {Treyer}, {Barlow}, {Forster}, {Friedman}, {Morrissey},
  {Neff}, {Small}, {Bianchi}, {Donas}, {Heckman}, {Lee}, {Madore}, {Milliard},
  {Rich}, {Szalay}, {Welsh}, \& {Yi}}]{2007wyder}
{Wyder}, T.~K., {Martin}, D.~C., {Schiminovich}, D., {et~al.} 2007, \apjs, 173,
  293, \dodoi{10.1086/521402}

\bibitem[{{Zaritsky} {et~al.}(1993){Zaritsky}, {Smith}, {Frenk}, \&
  {White}}]{1993Zaritsky}
{Zaritsky}, D., {Smith}, R., {Frenk}, C., \& {White}, S. D.~M. 1993, \apj, 405,
  464, \dodoi{10.1086/172379}

\bibitem[{{Zaritsky} {et~al.}(1997){Zaritsky}, {Smith}, {Frenk}, \&
  {White}}]{1997Zaritsky}
---. 1997, \apj, 478, 39, \dodoi{10.1086/303784}

\end{thebibliography}

\appendix
\section{Testing Resolution and Star-formation Tracers in Simulations}\label{sec:restests}
To test our results for convergence with resolution, we consider 5 high resolution (L1) volumes (10 hosts) from the APOSTLE simulations with $m_{\rm DM}\sim5\times10^4\,\msun$, $m_{\rm star}\sim1\times10^4\,\msun$ and 6 high resolution (Level 3) hosts from Auriga with $m_{\rm DM}\sim4\times10^4\,\msun$, $m_{\rm star}\sim6\times10^3\,\msun$.\ The satellites in these sets of simulations are treated as in $\S$\ref{subsec:sims}: subhalos are selected with SUBFIND, the \SAGAII{} spatial selection criteria are applied, and all physical properties are defined identically.\ 

We also test for any dependence on our star-formation metric, i.e.\ SFR estimated from the gas particles/cells.\ We repeat our quenched fraction estimates using the star-formation rate calculated based on the average number of star particles created over the last gigayear (SFR-1Gyr).\ This measure provides a more accurate estimate of a satellite's SFR compared to SFR dervied from star particles on shorter timescales that are susceptible to shot noise given the time and particle resolutions in the simulations.\ However, this measure will lead to a marginally higher number of star-forming satellites relative to our fiducial as it will include satellites that may have ceased forming stars within the last gigayear.\

\begin{figure*}[htb!]
\includegraphics[width=\textwidth]{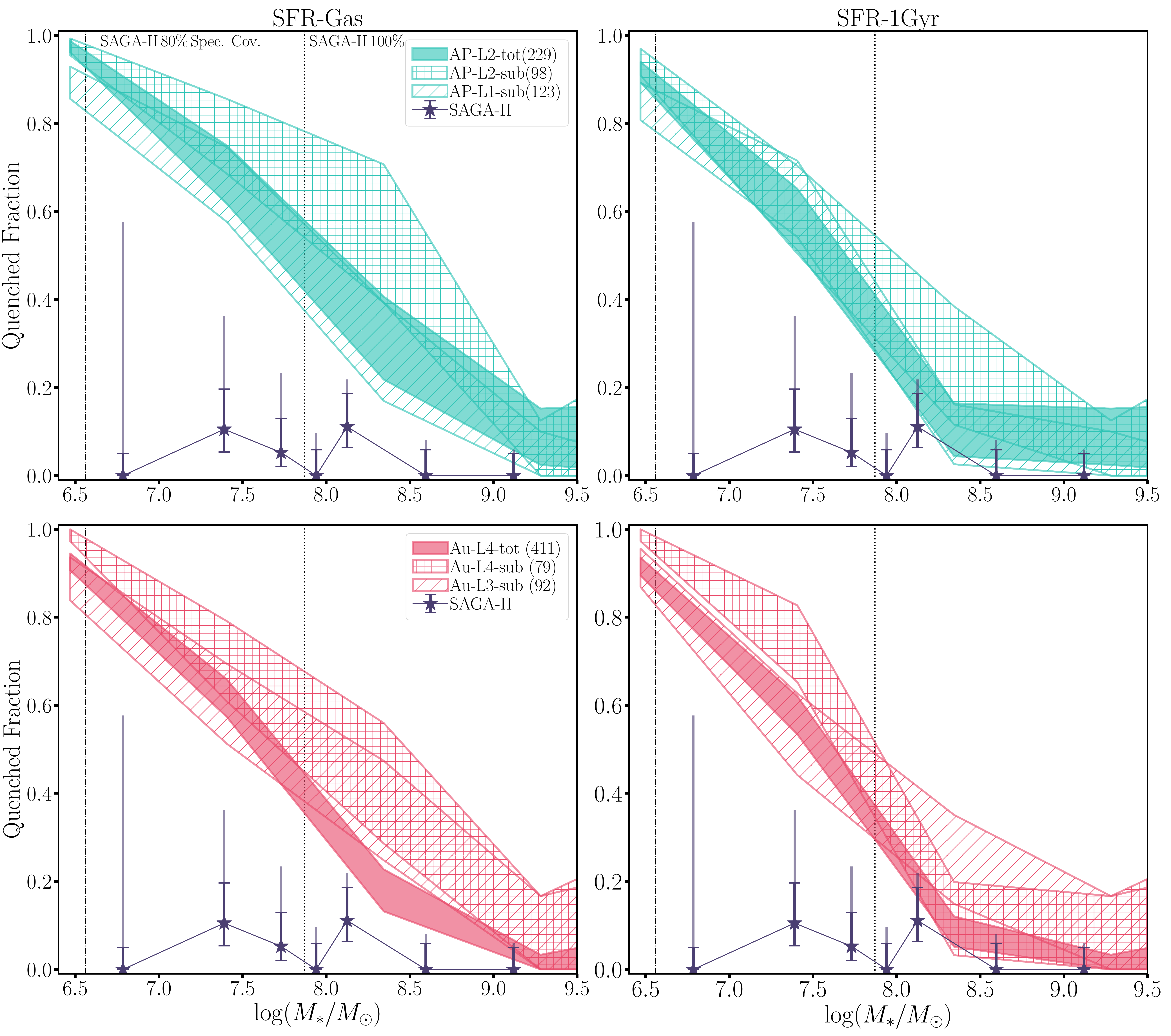}
\caption{Quenched fraction as a function of stellar mass plotted in the same manner as Figure \ref{fig:quenchfracsfhist}.\ The top and bottom rows show the APOSTLE and Auriga simulations, respectively.\ The left column plots the quenched fractions calculated using the SFRs derived from the gas, while the right column shows the quenched fractions calculated using the average SFR over the past 1 Gyr based on the stellar particles/cells.\ The filled bands show the 68$\%$ confidence intervals for total simulation samples at the standard resolution, the diagonal-hatched bands show the higher resolution subset, and the grid-hatched bands show the subset of standard resolution volumes that match the high resolution volumes.\ For reference, the total number of satellites used in these comparisons is listed in the legend.\ }
\label{fig:quenchedrestest}
\end{figure*}

The results of these tests are shown in Figure \ref{fig:quenchedrestest}.\ The left column shows the quenched fractions as a function of stellar mass for the APOSTLE (top) and Auriga (bottom) samples using our fiducial SFR definition, i.e.\ $\mathrm{SFR_{sim}=SFR_{gas}}>0\,\msun\,\mathrm{yr^{-1}}$.\ The shaded regions show the 68$\%$ confidence intervals and correspond to the total sample at the standard (filled; 229 APOSTLE satellites and 411 Auriga satellites), the subset simulated at higher resolution (diagonal hatched pattern; 123 APOSTLE satellites and 92 Auriga satellites), and the matching subset of volumes at the standard resolution (grid hatched pattern; 98 APOSTLE satellites and 79 Auriga satellites).\ The right column of Figure \ref{fig:quenchedrestest} plots the same samples except using the alternative SFR based on the star particles, i.e.\ $\mathrm{SFR_{1Gyr}}>0\,\msun\,\mathrm{yr^{-1}}$.\ In all 4 panels, we can see that the discrepancy in quenched fractions as a function of stellar mass, our primary result, remains between the observed and simulated samples at higher resolution and with an alternative star-formation definition.\

\end{document}